\documentclass[aps, pra, reprint, floatfix]{revtex4-1}
\usepackage[T1]{fontenc}            % Allows to insert special fonts
\usepackage[utf8]{inputenc}       	% Allows you to insert special characters (UTF8)
\usepackage[english]{babel}       	% Language used in the document.
\usepackage{microtype}              % allows LaTex to stretch text for better text justify

\usepackage{graphicx}             	% Allows to insert pictures
\usepackage[table]{xcolor}

\usepackage{amsmath}              	% Additional mathematics
\usepackage{physics}              	% Shortcuts for many physics related things.
\usepackage{siunitx}                % Display measurement data easily
\usepackage{bm}				        % bold math

\usepackage{dcolumn}			    % Align table columns on decimal point
\usepackage{hyperref}     		    % Hyperlinks within pdf and toward internet
\hypersetup{
	colorlinks = true,
	linkcolor=,
	citecolor=red,
	urlcolor=blue
}

\renewcommand*{\url}[1]{(url:~\href{\detokenize{#1}}{\detokenize{#1}})}

\newcommand{\y}{\cellcolor{yellow!50}}
\newcommand{\phm}{\phantom{-}}

\newcommand{\ee}{{\text e}}  % roman exponent
\newcommand{\ii}{{\text i}}  % roman complex i
\begin{document}
%%%%%%%%%%%%%%%%%%%%%%%%%%
%%%%%%%%%%%%%%%%%%%%%%%%%%
\title{Resolving features and derivatives in noisy data using weighted Whittaker--Henderson smoothing}
%%%%%%%%%%%%%%%%%%%%%%%%%%
\author{Bert Mulder}
    \email{l.mulder1@tue.nl} 
\author{Ad Lagendijk}
\author{Willem L. Vos}
    \email{w.l.vos@tue.nl} 
\affiliation{Complex Photonic Systems (\href{https://nano-cops.com/}{COPS}) Chair, Faculty of Science and Technology (S\&T), University of Twente (UT), P.O.~Box~217, 7500~AE Enschede, The Netherlands}
\affiliation{Complex Photonic Systems (\href{https://nano-cops.com/}{COPS}) Group, Photonic and Semiconductor Nanostructures (PSN) Chair, Department of Applied Physics and Science Education (APSE), Eindhoven University of Technology (TU/e), P.O.~Box~513, 5600~MB Eindhoven, The Netherlands}
%%%%%%%%%%%%%%%%%%%%%%%%%%
\date{8 June 2026}
%%%%%%%%%%%%%%%%%%%%%%%%%%
%%%%%%%%%%%%%%%%%%%%%%%%%%
\begin{abstract} % 227/250 words
A frequently occurring challenge in experimental and numerical observations is how to resolve features, such as spectral peaks - with center, width, height - and derivatives from measured data with unavoidable noise. 
Although many smoothing procedures exist, most are ineffective at reducing noise when the widths of the features varies strongly.
Therefore, we modify the Whittaker--Henderson smoothing procedure to locally balance the spectral features and the noise.
The central contribution of our procedure is that we introduce adjustable weights that are optimized using cross-validation. 
Using the measurement errors, a straightforward error analysis of the smoothed results is feasible. 
To illustrate the effectiveness of our smoothing algorithm, we derive for an optical Bragg reflector nanostructure the chirp of an optical pulse (group delay dispersion) using synthetic phase data with noise.
The smoother faithfully reconstructs the group delay dispersion, reducing noise by more than a factor \num{40}, allowing to identify details that otherwise remain buried in noise.
Finding the optimal weights using the limited-memory BFGS algorithm for $N=1000$ complex valued reflectivity data points takes on average less than two seconds on a typical computer.
To further illustrate the power of our smoother, we introduce a general framework to solve commonly occurring difficulties in data and data analysis; how to properly smoothen unequally sampled data, how to identify and quantify discontinuities, including discontinuous derivatives or kinks, how to properly smooth data in the vicinity of boundaries to the data domains, and multi-dimensional smoothing.
\end{abstract}
%%%%%%%%%%%%%%%%%%%%%%%%%%
%%%%%%%%%%%%%%%%%%%%%%%%%%
\maketitle
%%%%%%%%%%%%%%%%%%%%%%%%%%
%%%%%%%%%%%%%%%%%%%%%%%%%%%%%%%%%%%%%%%
\section{Introduction}\label{sec:introduction}
%%%%%%%%%%%%%%%%%%%%%%%%%%%%%%%%%%%%%%%
\par
\begin{figure}[t]
    \centering
    \includegraphics[width=.92\linewidth]{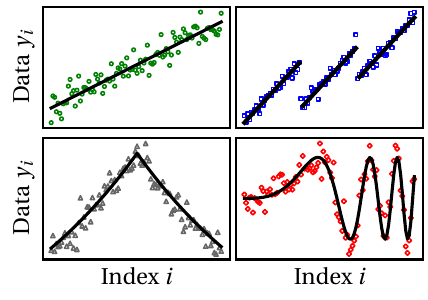}
    \caption{
        Schematic illustrations of discrete data $(i, y_i)$ with noise in which an observer may seek to determine various salient features. 
        Top left: noisy data from which one may want to obtain the slope.
        Top right: data with two jumps that one may wish to determine. 
        Bottom left: the challenge of identifying a cusp in data with noise. 
        Bottom right: noisy data with a range of feature widths one may wish to resolve. 
     }
     \label{fig:1st_figure}
\end{figure}

 %%%%%%%%%%%%%%%%%%%%%
A basic feature of natural sciences is the collection of sequences of observations~\cite{Sussman2016chapter, Shapere1982PoS}, where the observations inevitably suffer from various imperfections~\cite{Horowitz1983book, Goodman2000book}.
Many different kinds of such sequential observations are made in our field of research in optics, varying from spectral measurements where one records intensities, \textit{e.g.}, reflectivity, scattering and Raman, as a dependent variable as a function of a control variable like wavelength or optical frequency, angle or momentum, position, or time.
As a robust primary interpretation of the observations $y_i$ as a function of some index $i$, one may wish to determine characteristic features, such as a slope, a jump, a peak, or a resonance frequency, see figure~\ref{fig:1st_figure}.
%%%%%%%%%%%%%%%%%%%%%%%%%%
%%%%%%%%%%%%%%%%%%%%%%%%%%
\par %%%%%%%%%%%%%%%%%%%%%
In all interpretations of observational data, also outside optics, one has to properly deal with imperfections in the data that hamper interpretation.
Examples include noise in the observation due to intrinsic effects and imperfect apparatus, or unequal or even random sampling of the data as illustrated in figure~\ref{fig:1st_figure}.
Therefore, one needs an educated filtering of the data to emphasize the property one seeks~\cite{Orfanidis2007book, Press2007book}.
\par %%%%%%%%%%%%%%%%%%%%%
The basic challenge with filtering is that noise and intrinsic features are treated the same.
One may reduce noise by filtering high frequencies, but this leads to an underestimation of peaks and valleys in the data.
Another issue is that it is hard to do a proper error estimation of the filtered data~\cite{Altman2013book}.
\par %%%%%%%%%%%%%%%%%%%%%
Filtering procedures are distinguished into two classes, namely causal filters, and non-causal filters that are also known as smoothers.
Notable examples of causal filters are the Wiener filter~\cite{Wiener1949book, Einicke2012book} and Kalman filter~\cite{Orfanidis2007book}.
The Wiener filter is the first proposed filter that is designed using a statistical approach, as opposed to the usual frequency response approach.
Similarly, Kalman filters combine previous measurements and a mathematical model of the system under study to estimate the optimal next value.
Therefore, this method is the best causal linear filter for tracking the state of a dynamic system.
For many kinds of data such as optical spectra, however, the use of a causal filter is detrimental.
The position and shape of features change due to unavoidable (frequency-dependent) lag of causal filters, making these filters unsuitable for applications where phase preservation is critical.
\par %%%%%%%%%%%%%%%%%%%%%
The class of non-causal filters include local regression~\cite{Buja1989AoS, Yee2015book} and smoothing splines~\cite{Eilers1996SS, Boor2001book}.
Local regression procedures such as LOESS~\cite{Cleveland1988JotASA}, LOWESS~\cite{Cleveland1979JotASA} and the famous Savitzky--Golay filter~\cite{Savitzky1964AC} use a moving window over the data.
The local smooth value is found using regression analysis within the window.
A limiting factor to the smoothness is the size of the window.
A larger window allows for smoother curves but also leads to more problems in interpreting data at the boundaries of the domain.
The smoothing spline procedure splits the data into many parts at positions called knots.
Higher-order derivatives at these knots are discontinuous, which may lead to artifacts when we are interested in these derivatives.
Furthermore, the locations of the knots determine for a part the smoothed result, which leads to some ambiguity in the smoothed result.
\par %%%%%%%%%%%%%%%%%%%%%
Our method is based on Whittaker--Henderson smoothing, that was introduced in 1922~\cite{Whittaker1922PEMS, Henderson1924TASA}, but seems to have been overlooked in literature~\cite{Eilers2003AC, Biessy2025AB}.
The Whittaker--Henderson smoothening procedure, from now on called in shorthand Whittaker--Henderson smoother, does not require any additional assumptions about the data, as it is not based on a model of the system under study, but on a general notion of roughness. 
Whittaker--Henderson smoothers thus offer the liberty to apply to many different types of data, in particular those for which no underlying model or analytic theory is known.
A second favorable property of Whittaker--Henderson smoothers is that all data is smoothed together, meaning there is no moving window and there are no knots. 
The lack of a moving window leads to increased high-frequency noise suppression, which is beneficial for extracting features such as (higher-order) derivatives.
A third favorable property is that Whittaker--Henderson smoothers do not use a set of basis functions, making it a simpler and therefore computationally cheaper method.
\par %%%%%%%%%%%%%%%%%%%%%
Whittaker--Henderson smoothing is an $\ell^2$-regularization technique and can be directly used as an alternative for, \textit{e.g.}, local-regression and smoothing-spline procedures.
Our main contribution to Whittaker--Henderson smoothing is to introduce adjustable weights that are optimized to locally balance the spectral features and the noise.
This allows to construct the most likely and maximally smooth curve that tracks noisy data (see figure~\ref{fig:1st_figure}) with credible-interval estimation. 
We present a practical method to find the optimal weights using a cross-validation metric.
This method is highly beneficial for a wide range of domains, ranging from, \textit{e.g.}, statistics, signal processing, spectroscopy, to econometrics. 
In our research in optics and nanophotonics~\cite{Demtroder2002book, Novotny2012book}, our method allows to identify resonances (with main properties center, width, amplitude), slopes (dispersion) and slow light, or even higher-order slopes (group velocity dispersion), see, \textit{e.g.}, Refs.~\cite{Imhof1999PRL, Gersen2005PRL, Krauss2008NatPhot, Baba2008NatPhot}. 
\par %%%%%%%%%%%%%%%%%%%%%
Our article is organized as follows:
Section~\ref{sec:wh_smoothing} sets the general framework of Whittaker--Henderson smoothing. 
It introduces the underlying principles, explains the roughness measure and how to do error analysis.
Section~\ref{sec:methods} describes assumptions on noise and our numerical implementation of Whittaker--Henderson smoothing.
Section~\ref{sec:weight_wh_smoothing} extends Whittaker--Henderson smoothing by introducing adjustable weights $w_i$ for each respective data point $i$, and how to use cross-validation to find the optimal weights $w_i$.
Section~\ref{subsec:application_gdd} has the main example of this manuscript.
We show quantitatively how to optimally reconstruct the optical group delay dispersion from noisy phase data with sharp features.
Sections~\ref{sec:uneven_spaced} to \ref{sec:mult_dim_filtering} provide qualitative overview on applications of the same general framework for many common use-cases and pathologies, namely unequal sampling, jumps, cusps, boundaries, and multidimensional smoothing.
%%%%%%%%%%%%%%%%%%%%%%%%%%%%%%%%%%%%%%%
\section{Whittaker--Henderson smoothing}\label{sec:wh_smoothing}
%%%%%%%%%%%%%%%%%%%%%%%%%%%%%%%%%%%%%%%
\par %%%%%%%%%%%%%%%%%%%%%
To begin with, we possess a 1D-array of measured data including noise $\mathbf{y}$ that is the input of our procedure.
The smoothed output data array is $\mathbf{z}$.
The Whittaker--Henderson smoothing method relies on minimizing a functional $Q[\mathbf{z}]$, that is defined as
%%%%%%%%%%%%%%%%%%%%%%%%%%
\begin{equation}
    Q[\mathbf{z}] \equiv \norm{\mathbf{z-y}}^2 + \alpha\norm{\mathbf{Dz}}^2. 
    \label{eq:wh_q_functional}
\end{equation}
%%%%%%%%%%%%%%%%%%%%%%%%%%
Here the first term $\norm{\mathbf{z-y}}^2$ is the squared distance (using the $\ell^2$-norm) between the smoothed result $\mathbf{z}$ and the data $\mathbf{y}$.
The second term $\alpha\norm{\mathbf{Dz}}^2$ is a measure of the "roughness" of the smoothed result, where $\mathbf{D}$ is a matrix operating on $\mathbf{z}$ and the smoothness parameter $\alpha\ge\num{0}$ controls how strong the roughness is penalized.
Hence, Whittaker--Henderson smoothing is a special case of ridge regression~\cite{Polak2001MST}, also known as Tikhonov regularization~\cite{Engl2000book} or penalized least-squares.
While there is a strong similarity to compressed sensing~\cite{Candes2006CPAM, Caravaca2023IC}, compressed sensing uses $\ell^1$-norm penalties, however, leading to sparse models~\cite{Tibshirani1996JRSSSB}.
\par %%%%%%%%%%%%%%%%%%%%%
Minimizing $Q[\mathbf{z}]$ corresponds to balancing the smoothness of the resulting output $\mathbf{z}$ versus the difference between the data $\mathbf{y}$ and the result $\mathbf{z}$.
With $\alpha=\num{0}$, the functional $Q[\mathbf{z}]$ is only minimized when $\mathbf{z}=\mathbf{y}$, hence no smoothing is applied.
The larger $\alpha$ becomes, the smoother the result $\mathbf{z}$ becomes, and ultimately the output is so smoothened that all features have been removed.
Consequently, it is important to find the smoothness parameter $\alpha$ that gives the best compromise between smoothness and accuracy.
The optimal smoothness parameter $\alpha$ is usually found using one of the many available statistical tests.
\par %%%%%%%%%%%%%%%%%%%%%
Using the variational principle, we find that $\mathbf{z}$ minimizing equation~\ref{eq:wh_q_functional} is found by solving
%%%%%%%%%%%%%%%%%%%%%%%%%%
\begin{equation}
    \left(\mathbf{I} +\alpha\mathbf{D}^{\rm T} \mathbf{D}\right)\mathbf{z} = \mathbf{y}, 
    \label{eq:wh_linear_system}
\end{equation}
%%%%%%%%%%%%%%%%%%%%%%%%%%
where $\mathbf{I}$ is the identity matrix and $\mathbf{D}^{\rm T}$ the matrix transpose of $\mathbf{D}$.
Generally, this linear system consists of sparse matrices, scales linearly with the number of data points, and is thus very efficient to solve.
%%%%%%%%%%%%%%%%%%%%%%%%%%
\subsection{Understanding the Roughness Measure}\label{subsec:roughness_meas}
%%%%%%%%%%%%%%%%%%%%%%%%%%
\par %%%%%%%%%%%%%%%%%%%%%
The roughness measure is often taken to be a derivative of $\mathbf{z}$~\cite{Simonoff1996book}, with the matrix $\mathbf{D}$ acting as a finite difference operator.
In this paper we will assume $\mathbf{D}$ is always some linear combination of finite difference operators.
Other kinds of roughness measures can be used as well and can be tailored depending on the type of data.
Care has to be taken when picking a roughness measure, as an improper measure may exaggerate specific features, leading to undesirable artifacts.
\par %%%%%%%%%%%%%%%%%%%%%
Historically, three kinds of roughness measures are commonly used.
Bohlmann smoothers~\cite{Bohlmann1899NGWG}, introduced two decades before Whittaker--Henderson smoothers, use a first difference measure.
This measure can be understood as penalizing the length of the curve $\mathbf{z}$, thus, a shorter curve is favorable.
Hodrick–Prescott smoothers~\cite{Hodrick1997JMCB}, which are used in macroeconomics to remove the cyclical component from a time-series, use a second difference measure.
This can be understood as penalizing bending of the curve $\mathbf{z}$, thus, a flatter curve is favorable.
Whittaker--Henderson smoothers originally used a third difference measure.
This penalizes a change in the bending of the curve $\mathbf{z}$.
\par %%%%%%%%%%%%%%%%%%%%%
For every roughness measure $\mathbf{D}$ there exist null vectors $\mathbf{v}$ that fulfill the relation
%%%%%%%%%%%%%%%%%%%%%%%%%%
\begin{equation}
    \norm{\mathbf{D}\mathbf{v}}^2 \equiv 0. 
    \label{eq:insens_wh}
\end{equation}
%%%%%%%%%%%%%%%%%%%%%%%%%%
Since these curves have no roughness penalty, any data $\mathbf{y}$ will converge to a linear combination of these null vectors with sufficiently strong smoothing.
This means that for an $n^{\rm th}$ derivative roughness penalty, the smoothed result will converge to an $(n-1)^{\rm th}$-order polynomial.
%%%%%%%%%%%%%%%%%%%%%%%%%%
\subsection{Error Analysis}\label{sec:err_analysis}
%%%%%%%%%%%%%%%%%%%%%%%%%%
\par %%%%%%%%%%%%%%%%%%%%%
Smoothing reduces the number of effective degrees of freedom $d_{\rm eff}$.
We calculate the degrees of freedom of the smoothed result using the smoother (or influence) matrix $\mathbf{H}$~\cite{Wood2006book}, that is defined implicitly as
%%%%%%%%%%%%%%%%%%%%%%%%%%
\begin{equation}
    \mathbf{H y} \equiv \mathbf{z}. 
    \label{eq:wh_smoother_matrix}
\end{equation}
%%%%%%%%%%%%%%%%%%%%%%%%%%
It is called the smoother matrix, because it transforms the data $\mathbf{y}$ into the smoothed result $\mathbf{z}$.
This definition also holds for modified Whittaker--Henderson smoothers.
When strong smoothing is applied, the diagonal of $\mathbf{H}$ tends to zero, while in the case of no smoothing the diagonal of $\mathbf{H}$ will be equal to unity, \textit{i.e.}, $\mathbf{H}$ will then be equal to the identity matrix.
The effective number of degrees of freedom $d_{\rm eff}$ is defined as~\cite{Buja1989AoS}
%%%%%%%%%%%%%%%%%%%%%%%%%%
\begin{equation}
    d_{\rm eff}\equiv\text{Tr}\{\mathbf{H}\}. 
    \label{eq:dof_H}
\end{equation}
%%%%%%%%%%%%%%%%%%%%%%%%%%
In the absence of smoothing the number of degrees of freedom is equal to the number of data points $d_{\rm eff}=N$.
With smoothing the number of degrees of freedom reduces and tends to near zero for strong smoothing.
\par %%%%%%%%%%%%%%%%%%%%%
Smoothing introduces a bias to the error estimation, since the number of effective degrees of freedom is reduced.
A better estimate of the error is found by using the smoother matrix as defined in equation~\ref{eq:wh_smoother_matrix}.
Assuming we found the smooth curve $\mathbf{z}$ such that the residual $(\mathbf{y-z})$ is independent and normally distributed, we then use the result of Marra and Wood~\cite{Marra2012SJS} to estimate an unbiased credibility interval $\sigma_{\textbf{z}}$ for $\mathbf{z}$, where the credibility interval for each data point $i$ is given by
%%%%%%%%%%%%%%%%%%%%%%%%%%
\begin{equation}
    \sigma_{z,i} = \sigma_{y,i}\sqrt{{\rm H}_{ii}}, 
    \label{eq:wh_confidence}
\end{equation}
%%%%%%%%%%%%%%%%%%%%%%%%%%
where $\sigma_{\mathbf{y}}$ is a known measurement error.
This is equivalent to rescaling the error by the square root of the degrees of freedom of each data point.
\par %%%%%%%%%%%%%%%%%%%%%
Another way of estimating the error is using the covariance matrix~\cite{Wood2006book}.
Assuming that the errors in the data are uncorrelated, we construct the covariance matrix $\Sigma_{\mathbf{z}}$ as
%%%%%%%%%%%%%%%%%%%%%%%%%%
\begin{equation}
    \mathbf{\Sigma}_{\mathbf z} = \mathbf{H}\,\text{diag}\{\sigma^2_{\mathbf y}\}\,\mathbf{H}^{\rm T}. 
    \label{eq:wh_covariance}
\end{equation}
%%%%%%%%%%%%%%%%%%%%%%%%%%
The credibility interval is then given by 
%%%%%%%%%%%%%%%%%%%%%%%%%%
\begin{equation}
    \sigma_{z,i} = \sqrt{\Sigma_{z,ii}}. 
    \label{eq:wh_conf_from_covariance}
\end{equation}
%%%%%%%%%%%%%%%%%%%%%%%%%%
We find that the credibility intervals calculated using equation~\ref{eq:wh_confidence} and equation~\ref{eq:wh_conf_from_covariance} are in very good mutual agreement.
Therefore, we choose to calculate the credibility interval using equation~\ref{eq:wh_conf_from_covariance}, as this requires fewer assumptions.
%%%%%%%%%%%%%%%%%%%%%%%%%%
\begin{figure}[t]
    \centering
    \includegraphics[width=\linewidth]{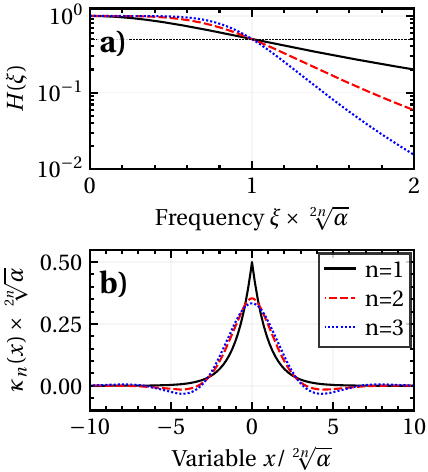}
    \caption{
        \textbf{(a)} Frequency response for the first three orders of continuous Whittaker--Henderson smoothers.
        The dotted line indicates where the amplitude of $H(\xi)$ has decreased to one half.
        Higher-order filters have a flatter passband and their high-frequency response decays faster.
        \textbf{(b)} Corresponding convolutions kernels $\kappa_n(x)$ for the first three orders of continuous Whittaker--Henderson filters.
        The kernels are symmetric and decay exponentially on both sides of the origin.
        Higher-order filters show more ringing.
        }
    \label{fig:wh_freq_resp}
\end{figure}
%%%%%%%%%%%%%%%%%%%%%%%%%%
%%%%%%%%%%%%%%%%%%%%%%%%%%
\subsection{Continuous Analytical Frequency Response}\label{subsec:freq_resp}
%%%%%%%%%%%%%%%%%%%%%%%%%%
\par %%%%%%%%%%%%%%%%%%%%%
To better understand the smoothing properties of the Whittaker--Henderson smoother, we approximate the \emph{discrete} data $y_i$ with a \emph{continuous} function $y(x)$.
After this excursion to the continuous case, we return to discrete smoothers.
The generalization to continuous functions allows us to study the frequency response of the smoother analytically.
This approximates the case of smoothing densely sampled data far from the boundaries of the data domain.
For the continuous case we have to find the output function $z(x)$ that minimizes a functional $Q_{\rm c}[z(x)]$ defined as
%%%%%%%%%%%%%%%%%%%%%%%%%%
\begin{equation}
    \begin{split}
        Q_{\rm c}[z(x)] \equiv \int_{-\infty}^{\infty} \vert z(x)-y(x)\vert^2 dx \\
        + \alpha\int_{-\infty}^{\infty} \vert\hat{D}z(x)\vert^2 dx, 
    \end{split}
    \label{eq:cont_Q_func}
\end{equation}
%%%%%%%%%%%%%%%%%%%%%%%%%%
with $\hat{D}$ a linear differential operator as roughness measure and $y(x)$ the input function to be smoothed.
Assuming that $y(x)$ vanishes as $x\rightarrow\pm\infty$, then the function $z(x)$ that minimizes $Q_{\rm c}[z(x)]$ is found by solving the differential equation
%%%%%%%%%%%%%%%%%%%%%%%%%%
\begin{equation}
    z(x) + \alpha\hat{D}^\dagger\hat{D}z(x) = y(x), 
    \label{eq:diff_eqq_wh}
\end{equation}
%%%%%%%%%%%%%%%%%%%%%%%%%%
where the Hermitian conjugate of a derivative is given by
%%%%%%%%%%%%%%%%%%%%%%%%%%
\begin{equation}
    \left[\left(\frac{\dd}{\dd{x}}\right)^n\right]^\dagger = \left(-\frac{\dd}{\dd{x}}\right)^n. 
    \label{eq:hermitian_deriv}
\end{equation}
%%%%%%%%%%%%%%%%%%%%%%%%%%
The linear differential equation is readily solved in the frequency-domain using a Fourier transform.
We find that the frequency response $H(\xi)$ of the filter is 
%%%%%%%%%%%%%%%%%%%%%%%%%%
\begin{equation}
    H(\xi)\equiv \frac{\tilde{z}(\xi)}{\tilde{y}(\xi)} = \frac{1}{1 + \alpha\vert\tilde{D}(\xi)\vert^2}, 
    \label{eq:fourier_wh}
\end{equation}
%%%%%%%%%%%%%%%%%%%%%%%%%%
where $\tilde{D}(\xi)$ is the linear differential operator $\hat{D}$ in the frequency-domain.
The transformation of a differential operator to the frequency-domain is 
%%%%%%%%%%%%%%%%%%%%%%%%%%
\begin{equation}
    \left(\frac{\dd}{\dd{x}}\right)^n \xrightarrow{\mathcal{F}} (\ii\xi)^n. 
    \label{eq:fourier_deriv}
\end{equation}
%%%%%%%%%%%%%%%%%%%%%%%%%%
The result in equation~\ref{eq:fourier_wh} has three important implications.
Firstly, we find that the frequency response is always real and positive, meaning it is a zero-lag filter.
Secondly, if the differential operator $\hat{D}$ is of order $n$, then we find that the amplitude of the frequency response will decay as $\xi^{-2n}$ for high frequencies. 
Thirdly, we find that a higher-order differential operator leads to a flatter passband.
The frequency response $H(\xi)$ for the first three orders of Whittaker--Henderson smoothers are shown in figure~\ref{fig:wh_freq_resp}\,(a).
We observe that all but the first-order ($n=1$) smoother have a flat passband.
As the order $n$ increases the passband becomes flatter and high frequencies become increasingly suppressed.
\par %%%%%%%%%%%%%%%%%%%%%
The solution $z(x)$ of equation \ref{eq:diff_eqq_wh} can be written as a convolution of a kernel $\kappa(x)$ with the input function $y(x)$ as
%%%%%%%%%%%%%%%%%%%%%%%%%%
\begin{equation}
    z(x) = \int_{-\infty}^{\infty}{y(x') \kappa(x-x') dx'}. 
    \label{eq:convolution_define}
\end{equation}
%%%%%%%%%%%%%%%%%%%%%%%%%%
The first-order ($n=1$) Whittaker--Henderson smoother is equivalent to smoothing with the (exponential) Poisson window function~\cite{Smith2011book}
%%%%%%%%%%%%%%%%%%%%%%%%%%
\begin{equation}
    \kappa_1(x) = \frac{1}{2\sqrt{\alpha}}\ee^{-\vert x\vert/\sqrt{\alpha}}, 
    \label{eq:kernel_k1}
\end{equation}
%%%%%%%%%%%%%%%%%%%%%%%%%%
and is shown in figure~\ref{fig:wh_freq_resp}\,(b).
Due to the absolute value, the kernel has a cusp at $x=0$.
The second- ($n=2$) and third- ($n=3$) order kernels $\kappa_n(x)$ are also illustrated in figure~\ref{fig:wh_freq_resp}\,(b).
The kernel function for the second-order case is similar to the basis function of a cubic regression spline smoother~\cite{Silverman1985JRSSB}. 
Higher-order convolution kernels become increasingly complex.
The expressions for the convolution kernels $\kappa_n(x)$ of the second-, third- and fourth-order Whittaker--Henderson smoothers are
%%%%%%%%%%%%%%%%%%%%%%%%%%
\begin{equation}
    \kappa_2(x) = \frac{1}{2\sqrt[4]{\alpha}}\sin\left(\frac{\vert x \vert}{\sqrt[4]{4\alpha}} +\frac{\pi}{4}\right) \ee^{-\vert x\vert/\sqrt[4]{4\alpha}}, 
    \label{eq:kernel_k2}
\end{equation}
%%%%%%%%%%%%%%%%%%%%%%%%%%
%%%%%%%%%%%%%%%%%%%%%%%%%%
\begin{multline}
    \kappa_3(x) = \frac{1}{3\sqrt[6]{\alpha}}\Bigg[\frac{1}{2} \ee^{-\vert x\vert/2\sqrt[6]{\alpha}} \\
    +\sin{\left(\frac{\sqrt{3}\vert x \vert}{2\sqrt[6]{\alpha}}+\frac{\pi}{6}\right)}\Bigg] \ee^{-\vert x\vert/2\sqrt[6]{\alpha}},
    \label{eq:kernel_k3}
\end{multline}
%%%%%%%%%%%%%%%%%%%%%%%%%%
%%%%%%%%%%%%%%%%%%%%%%%%%%
\begin{multline}
    \kappa_4(x) = \frac{1}{4\sqrt[8]{\alpha}}\Bigg[ \\
    \hspace{1.52em}\ee^{-\frac{\vert x\vert}{\sqrt[8]{\alpha}}\cos{(\frac{\pi}{8})}}\cos{\left(\frac{\vert x\vert}{\sqrt[8]{\alpha}}\sin{\left(\frac{\pi}{8}\right)} -\frac{\pi}{8} \right)}\\ 
    +\ee^{-\frac{\vert x\vert}{\sqrt[8]{\alpha}}\sin{(\frac{\pi}{8})}}\sin{\left(\frac{\vert x\vert}{\sqrt[8]{\alpha}}\cos{\left(\frac{\pi}{8}\right)} + \frac{\pi}{8} \right)} \Bigg]. 
    \label{eq:kernel_k4}
\end{multline}
%%%%%%%%%%%%%%%%%%%%%%%%%%
All Whittaker--Henderson convolution kernels $\kappa_n(x)$ are symmetric, decay exponentially on both sides of the origin, and have a discontinuity at $x=0$ due to the absolute value.
%%%%%%%%%%%%%%%%%%%%%%%%%%%%%%%%%%%%%%%
\section{Methods}\label{sec:methods}
%%%%%%%%%%%%%%%%%%%%%%%%%%%%%%%%%%%%%%%
%%%%%%%%%%%%%%%%%%%%%%%%%%
\subsection{Assumptions on Noise}\label{subsec:noise_assumptions}
%%%%%%%%%%%%%%%%%%%%%%%%%%
When trying to remove noise from measured data, it is important to know the statistical properties of the noise. 
In this article we always assume that noise is independent, homoscedastic, \textit{i.e.}, when all random variables have the same variance, normally distributed and stationary.
Our assumptions represent the optimal case for linear (penalized) least-squares smoothers, which does not necessarily hold for real-world data.
However, even if these assumptions are not exactly valid, Whittaker--Henderson smoothing can still be successfully applied with some limitations.
%%%%%%%%%%%%%%%%%%%%%%%%%%
\subsection{Numerical Implementation}\label{subsec:implementations}
%%%%%%%%%%%%%%%%%%%%%%%%%%
\par %%%%%%%%%%%%%%%%%%%%%
All our results are calculated using Python scripts running on a standard Lenovo ThinkPad laptop with a 13\textsuperscript{th} Gen Intel(R) Core(TM) i7-13700H processor (\SI{2.40}{\giga\hertz} base, \SI{5.00}{\giga\hertz} boost). 
To model noise, we use a pseudo-random noise generator as implemented by the "normal"~function from the {NumPy~random} module.
For reproducibility, the random number generator is initialized with a seed.
Sparse linear systems are solved using the {SciPy~sparse} module.
Initializing the sparse matrices of the linear system in equation~\ref{eq:wh_linear_system} to smooth $N=\num{e5}$ data points takes on average about \SI{13}{\milli\s}.
Solving this linear system then takes on average about \SI{21}{\milli\s}.
The numerical complexity is unknown to us, but appears linear for large $N$.
Different and potentially faster methods are available in literature~\cite{Weinert2007CSDA}.
We find the optimal smoothing parameter $\alpha$, or any other single filtering parameter, such as moving-window widths, using the generalized cross-validation method~\cite{Golub1979T, Lukas2006IP}.
All Python code is available as per \hyperref[sec:DAS]{data availability statement}.
%%%%%%%%%%%%%%%%%%%%%%%%%%%%%%%%%%%%%%%
\section{Weighted Whittaker--Henderson smoothing}\label{sec:weight_wh_smoothing}
%%%%%%%%%%%%%%%%%%%%%%%%%%%%%%%%%%%%%%%
%%%%%%%%%%%%%%%%%%%%%%%%%%
\subsection{Preliminaries}\label{subsec:weight_preliminary}
%%%%%%%%%%%%%%%%%%%%%%%%%%
A modification of Whittaker--Henderson smoothing is by adding weights $w_i$ for each data point~\cite{Eilers2003AC,Nocon2012SAJ,Biessy2025AB}.
We have to minimize a functional $Q_{\rm w}[\mathbf{z}]$ defined as
%%%%%%%%%%%%%%%%%%%%%%%%%%
\begin{equation}
    Q_{\rm w}[\mathbf{z}] \equiv \sum_{i=1}^N {w_i \abs{y_i - z_i}^2} 
    + \alpha \sum_{i=1}^N {\abs{({\rm D}z)_i}^2}. 
    \label{eq:weighted_Q}
\end{equation}
%%%%%%%%%%%%%%%%%%%%%%%%%%
We find that $\mathbf{z}$ minimizing equation~\ref{eq:weighted_Q} is found by solving
%%%%%%%%%%%%%%%%%%%%%%%%%%
\begin{equation}
    \left(\mathbf{W} + \alpha\mathbf{D}^{\rm T} \mathbf{D}\right)\mathbf{z} = \mathbf{Wy}, 
    \label{eq:wh_weighted_linear_system}
\end{equation}
%%%%%%%%%%%%%%%%%%%%%%%%%%
with $\mathbf{W}$ a diagonal matrix consisting of the weights $w_i$ for all data points, defined as
%%%%%%%%%%%%%%%%%%%%%%%%%%
\begin{equation}
    \mathbf{W} \equiv \text{diag}\{w_1,w_2,\dots,w_N\}. 
    \label{eq:wh_weights}
\end{equation}
%%%%%%%%%%%%%%%%%%%%%%%%%%
A high weight $w_i$ causes a high penalty to the least-squares difference, so that locally $\mathbf{z}$ closely tracks the data $\mathbf{y}$.
With a weight set to zero ($w_i=0$), the corresponding data point $y_i$ will be completely ignored and smoothly interpolated instead.
This feature is useful for incomplete data sets, where data points are missing, or to exclude outliers.
Weighted Whittaker--Henderson smoothing is normally used whenever there are unequal variances in observations, for example when noise is heteroscedastic or non-stationary. 
The weights $w_i$ are then proportional to the inverse noise variances.
We use the weights to apply different levels of smoothing to different parts of the data, which is vital to optimally resolve complex features in noisy data.
%%%%%%%%%%%%%%%%%%%%%%%%%%
\subsection{Find the Optimal Weights by Cross-Validation}\label{subsec:find_weights}
%%%%%%%%%%%%%%%%%%%%%%%%%%
\par %%%%%%%%%%%%%%%%%%%%%
For given input data $\mathbf{y}$, we want to find the optimal weight-to-smoothness ratio $w_i/\alpha$ such that the smoothed output $\mathbf{z}$ is \emph{maximally smooth} while the \emph{residual} $(\mathbf{y-z})$ \emph{remains random}.
Previous approaches include the balancing of the trade-off between bias and variance~\cite{Nowak1997IEEESPL}, iterative procedures using Gaussian noise statistics~\cite{Urbas2011AS}, or local and non-local sample correlations to remove noise~\cite{Li2021CSSP}. 
In contrast, we find the optimal weights directly using a statistical cross-validation method. 
\par %%%%%%%%%%%%%%%%%%%%%
Cross-validation is a statistical tool to evaluate how well the smoothed data correlates with the noisy data, by using a subset of the noisy data to predict the rest of the smoothed data~\cite{Wood2006book}.
The cross-validation value is the RMS difference between the predicted smooth output data and the real noisy input data.
In the case of weighted Whittaker--Henderson smoothing, a data point $i$ is predicted by setting its respective weight $w_i$ to zero.
A smaller cross-validation value means a better prediction, indicating a better matching fit between the smoothed data and the noisy data, and \textit{vice versa} a large cross-validation value indicates a worse fit.
By minimizing the cross-validation metric we thus construct the most likely smooth curve that tracks our data.
\par %%%%%%%%%%%%%%%%%%%%%
Usual leave-one-out cross-validation~\cite{Simonoff1996book} is extremely computationally expensive as it requires as many smooths as there are data points to calculate a single cross-validation value.
The minimization process simply takes much too long.
Therefore, we developed our own method where we split the data into two parts.
This way we only require two smoothing operations to calculate a single cross-validation value, which makes the minimization process computationally feasible.
Our method is similar to $k$-fold cross-validation with $k=2$.
However, in our case we assume a strong serial correlation in the ground truth and no serial correlation in the noise. 
Furthermore, random partitioning adds some variability, making minimization of the cross-validation metric much harder. 
Therefore, we use odd-even splitting of the data as opposed to the random partitioning used in $k$-fold cross-validation.
In the case that there is some serial correlation in the noise, it is necessary to split the data into more than two parts to break the serial correlation.
\par %%%%%%%%%%%%%%%%%%%%% 
For the first part of our method, we set even indexed $i=2,4,6,\dots$ of $w_i$ to zero to interpolate the even indexed $i=2,4,6,\dots$ of $\mathbf{z}_{\rm e}$ using the odd indexed $i=1,3,5,\dots$ of the data.
For the second part, we set odd indexed $i=1,3,5,\dots$ of $w_i$ to zero to interpolate the odd indexed $i=1,3,5,\dots$ of $\mathbf{z}_{\rm o}$ using the even indexed $i=2,4,6,\dots$ of the data.
We use both the even and odd indexed part of the data to make sure that we do not skip important features.
Our cross-validation metric is then
%%%%%%%%%%%%%%%%%%%%%%%%%%
\begin{equation}
    S_{\rm cv} \equiv \sqrt{\frac{1}{N}\sum_{i=1}^{N}
    \begin{cases}
        \abs{y_{i} - z_{{\rm e},i}}^2 & \hspace{-.1cm}\text{for even $i$,}\\
        \abs{y_{i} - z_{{\rm o},i}}^2 & \hspace{-.1cm}\text{for odd $i$.}
    \end{cases}}.
    \label{eq:s_cv}
\end{equation}
%%%%%%%%%%%%%%%%%%%%%%%%%%
In case of complex valued data, the real and imaginary components are taken jointly using the absolute difference.
\par %%%%%%%%%%%%%%%%%%%%%
One might be tempted to try to find the weights $w_i$ directly using a multivariate minimization algorithm based on our cross-validation metric.
This works up to a point but often noise in the data prevents us from finding the weights $w_i$ directly.
To get around this limitation we have to reduce the effects of noise by strongly reducing the number of degrees of freedom of the weights.
We do this by approximating the weights $w_i$ with a cubic b-spline.
The locations of the knots for the b-spline are found using the knot-vector constructor from the FORTRAN FITPACK package on the measured data.
The knot-vector constructor uses the least-squares criterion to find the best locations and the required number of knots.
We found that less knots also suffice and therefore we always pick about half the number of knots that the least-squares criterion gives.
The optimum weights $w_i$ are relatively insensitive to the exact number of knots.
In case of complex valued data, we find the knot locations for the real and imaginary part of the data separately and then combine the locations into a single knot-vector.
\par %%%%%%%%%%%%%%%%%%%%%
The optimum b-spline coefficients are found by minimizing our cross-validation metric using the multivariate bounded limited-memory BFGS algorithm as implemented by the "minimize" function of the {SciPy~optimize} module.
The L-BFGS-B algorithm is a quasi-Newton method that iteratively refines an initial estimate until a termination condition is reached.
In our case the termination condition is reached when all the components of the projected gradient are smaller than \num{e-6}.
%%%%%%%%%%%%%%%%%%%%%%%%%%%%%%%%%%%%%%%
\section{Extracting Derivatives from Noisy Data: Optical Group Delay Dispersion}\label{subsec:application_gdd}
%%%%%%%%%%%%%%%%%%%%%%%%%%%%%%%%%%%%%%%
\par %%%%%%%%%%%%%%%%%%%%%
The original motivation for our work is the study of the speed of light in advanced photonic nanostructures~\cite{Imhof1999PRL,Gersen2005PRL, Krauss2008NatPhot}, but the methods described in this section are universally applicable.
In our research we are interested in the group delay and the group delay dispersion that effectively correspond to taking first and second derivatives with optical frequency of spectral data.
Simply taking the numerical derivative of the measured data as finite differences will greatly amplify the effects of noise~\cite{Gosteva2005JOSAB}.
Therefore, a filtering method is required to increase the signal to noise ratio.
\par %%%%%%%%%%%%%%%%%%%%%
A practical example where smoothing is important is when determining the group delay dispersion from a widely-used distributed Bragg reflector of light~\cite{Fowles1989book, Euser2009RSI, Yuce2016OE}.
These reflectors are made of alternating layers of high- and low-index materials.
When the reflections of light from the different layers are all exactly in phase in the backward direction, then the reflected waves add up constructively, meaning that many small reflections add up to one large reflection.
Bragg reflectors made from many layers of lossless materials reach significantly higher reflectivities than metal reflectors, making them the logical choice when any amount of light loss would be detrimental.
Examples of applications include (laser) cavities, gravitational wave detectors~\cite{Granata2020CQG} and X-ray mirrors~\cite{Singhapong2024AMT} for EUV lithography and synchrotrons.
\par %%%%%%%%%%%%%%%%%%%%%
Group delay dispersion adds a chirp to a light pulse~\cite{Born1986book,Siegman1986Book,Jackson1998book,Boyd2008book}.
A positive chirp ($D_{2} > 0$), also known as normal dispersion, means that waves with a lower optical frequency (longer wavelength) arrive before waves with a higher optical frequency (shorter wavelength).
Conversely, a negative chirp ($D_{2} < 0$), also known as anomalous dispersion, means that waves with a higher optical frequency (shorter wavelength) arrive before waves with a lower optical frequency (longer wavelength).
In both cases the pulse duration increases and the peak power decreases, leading to less efficient nonlinear optical interactions and reduced time resolution in pump-probe experiments.
Understanding and compensating dispersion is therefore vital in any optical device or application employing nonlinear interactions or ultrafast processes.
\par %%%%%%%%%%%%%%%%%%%%%
The group delay dispersion $D_2(\omega)$ is defined as the second optical angular-frequency derivative of the optical phase~\cite{Diels2006Book,Pickering2021chapter},
%%%%%%%%%%%%%%%%%%%%%%%%%%
\begin{equation}
    D_{2}(\omega) \equiv \frac{\dd[2]{\phi(\omega)}}{\dd{\omega^2}},
    \label{eq:gdd}
\end{equation}
%%%%%%%%%%%%%%%%%%%%%%%%%%
where $\omega$ is the optical angular frequency and $\phi(\omega)$ is the measured optical frequency-dependent phase shift of light detected in a reflection experiment. 
The second derivative is extremely sensitive to noise, even small amounts of noise in the measured signal will completely obscure the calculated dispersion.
Generally, two methods are used to get around this problem.
For simple samples, such as homogeneous glass slabs with weak dispersion, the measured phase is often approximated by a high-order polynomial~\cite{Diddams1996JOSAB}.
The group delay dispersion is then calculated by differentiating the polynomial.
For more complex nanophotonic samples, however, where the phase shift is not readily described by a polynomial, the phase data is often filtered using, \textit{e.g.}, a Savitzky--Golay filter~\cite{Savitzky1964AC, Naganuma1990OL, Schmid2022MSAu} before it is differentiated.
\par %%%%%%%%%%%%%%%%%%%%%
As an alternative method to the optical angular-frequency derivative of the phase in equation~\ref{eq:gdd}, we obtain the dispersion $D_2$ from the optical angular-frequency derivatives of the complex reflectivity $r(\omega)$ as
%%%%%%%%%%%%%%%%%%%%%%%%%%
\begin{equation}
    D_2(\omega) = \Im\left\{\frac{1}{r(\omega)}\frac{\dd[2]{r(\omega)}}{\dd{\omega^2}} - \left(\frac{1}{r(\omega)}\frac{\dd{r(\omega)}}{\dd{\omega}}\right)^2\right\}. 
    \label{eq:gdd_from_amplitude}
\end{equation}
%%%%%%%%%%%%%%%%%%%%%%%%%%
Using our method to find the weights, described in section~\ref{subsec:find_weights}, we estimate the most likely smooth curve from our complex reflectivity data.
By finding the optimal weights $w_i$ for all data points we eliminate noise, while preserving true features.
From the smooth curve we calculate the derivatives $\partial_{\omega}r(\omega)$, and $\partial^2_{\omega}r(\omega)$, and then use equation~\ref{eq:gdd_from_amplitude} to calculate the group delay dispersion.
This is important, because the noise in real complex reflectivity data is often very nearly normally distributed.
In contrast, noise in the phase data is not normally distributed, which violates the assumptions made while deriving the method.
A second reason for using equation~\ref{eq:gdd_from_amplitude} is that the complex reflectivity is expected to be more or less continuous and smooth in physically realistic systems, while the group delay dispersion is not.
A diverging group delay dispersion is associated with large jumps in the phase; this would otherwise be much harder to properly reconstruct using a smoothing algorithm.
%%%%%%%%%%%%%%%%%%%%%%%%%%
\begin{figure}[t]
    \centering
    \includegraphics[width=\linewidth]{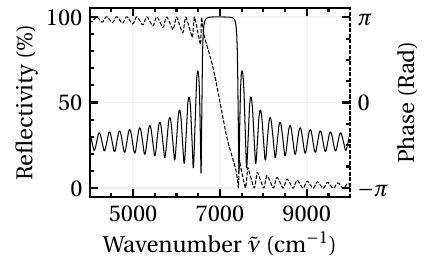}
    \caption{
        Intensity reflectivity (solid line) and reflection phase shift (dashed line) from a 30 pair GaAs/AlAs $\lambda/4$-layer Bragg reflector with a center wavenumber $\tilde{\nu}_{\rm c}=\SI{7000}{\per\cm}$ ($\lambda_{\rm c}=\SI{1429}{\nm}$).
        Data synthesized using transfer-matrix method~\cite{Yeh1977JOSA}.
    }
    \label{fig:model_data}
\end{figure}
%%%%%%%%%%%%%%%%%%%%%%%%%%
%%%%%%%%%%%%%%%%%%%%%%%%%%
\begin{figure}[t]
    \centering
    \includegraphics[width=\linewidth]{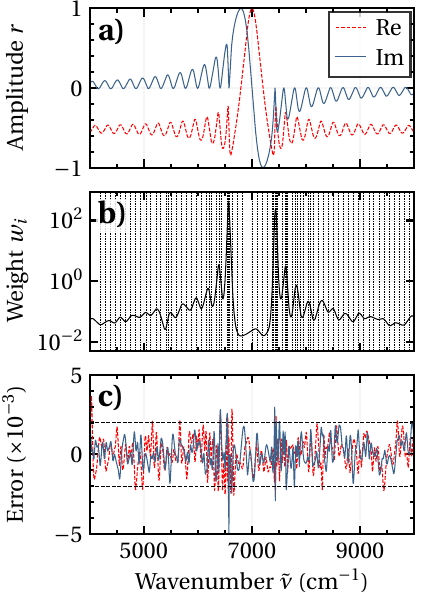}
    \caption{
        \textbf{(a)} Synthetic complex reflectivity data with noise.
        \textbf{(b)} The resulting weights $w_i$ to optimally smooth the complex reflectivity data.
        The domain is divided in 80 knot locations (dotted lines), the weights in between the knots are interpolated using a cubic b-spline.
        The smoothness parameter is fixed to $\alpha=\num{1}$.
        \textbf{(c)} Difference between smoothed synthetic data and the ground truth.
        The error is mostly smaller than the standard deviation (dashed lines) of the added noise.
    }
    \label{fig:wae_comparison}
\end{figure}
%%%%%%%%%%%%%%%%%%%%%%%%%%
%%%%%%%%%%%%%%%%%%%%%%%%%%
\begin{figure*}[t]
    \centering
    \includegraphics[width=\linewidth]{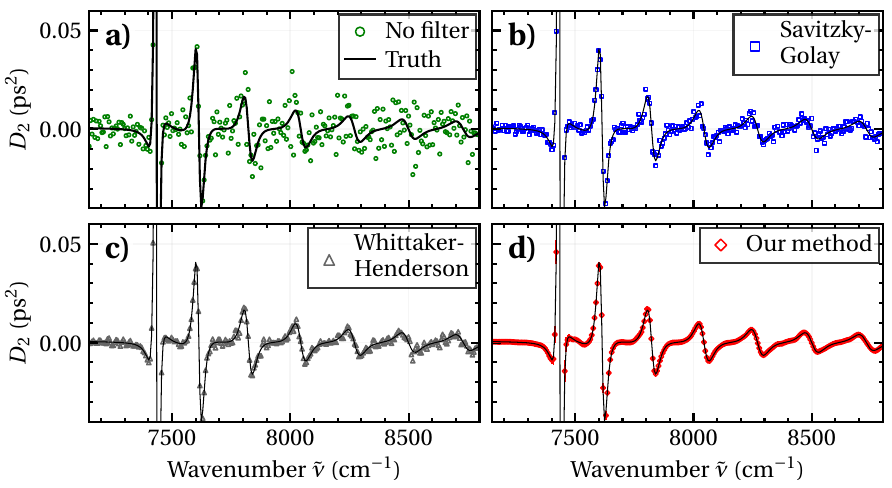}
    \caption{
        Calculated group delay dispersion $D_2(\omega)$ from noisy, synthetic reflectivity data of a Bragg reflector.
        Solid black traces are the true dispersion without noise.
        \textbf{(a)} The small amount of noise in the reflectivity data is enough to obscure all but the strongest dispersion features.
        \textbf{(b)} The noisy reflectivity data is filtered by a fourth-order Savitzky--Golay filter with an optimal window length of nine points, before the dispersion is calculated.
        The amount of noise is reduced, and some features become visible.
        \textbf{(c)} The noisy reflectivity data is smoothed by a fourth-order Whittaker--Henderson smoother with optimal smoothing parameter $\alpha=\num{0.110}$, before the dispersion is calculated.
        The Whittaker--Henderson smoother is better at reducing noise, while preserving high-frequency features, than the aforementioned Savitzky--Golay filter.
        \textbf{(d)} The noisy reflectivity data is reconstructed by our modified Whittaker--Henderson smoother, before the dispersion is calculated.
        The continuously variable smoothing weights allows strong noise reduction and faithful reproduction of the dispersion.
        The error bars denote the estimated $2\sigma_{\mathbf{z}}$ credibility interval.
    }
    \label{fig:method_gdd_comparison}
\end{figure*}
%%%%%%%%%%%%%%%%%%%%%%%%%%
%%%%%%%%%%%%%%%%%%%%%%%%%%
\begin{figure}[t]
    \centering
    \includegraphics[width=\linewidth]{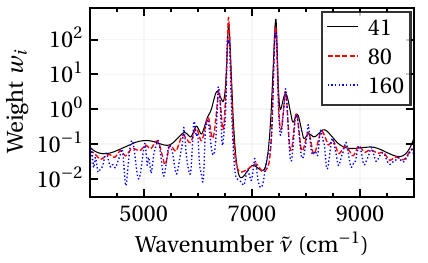}
    \caption{
        Comparison of the resulting optimal weights $w_i$ for 41, 80 and 160~knots respectively.
        The smoothness parameter is fixed to $\alpha=\num{1}$.
    }
    \label{fig:knots_comparison}
\end{figure}
%%%%%%%%%%%%%%%%%%%%%%%%%%
\par %%%%%%%%%%%%%%%%%%%%%
To obtain synthetic data on the complex reflectivity of a Bragg reflector, we use the transfer-matrix method~\cite{Yeh1977JOSA} for normally incident light.
As a realistic sample we use a Bragg reflector made from 30 pairs of GaAs/AlAs $\lambda/4$-layers stacked on a GaAs wafer in air.
For GaAs we assume a relative permittivity of $\varepsilon_{\rm GaAs}=11.6$ and for AlAs we assume a relative permittivity of $\varepsilon_{\rm AlAs}=8.44$.
The GaAs layer has a thickness of about $L_{\rm GaAs}=\SI{104.9}{\nm}$ and the AlAs layer a thickness of $L_{\rm AlAs}=\SI{122.9}{\nm}$.
The reflectivity spectrum is calculated from wavenumber $\tilde\nu=\SIrange{4000}{10000}{\per\cm}$ with step size $\delta\tilde\nu=\SI{6}{\per\cm}$.
The reflectivity and reflection phase shift are shown in figure~\ref{fig:model_data}, and reveal a broad photonic stopband with a relative bandwidth $\Delta\tilde{\nu}/\tilde{\nu}_{\rm c}= \SI{11.9}{\percent}$ centered at $\tilde{\nu}_{\rm c}=\SI{7000}{\per\cm}$ ($\lambda_{\rm c}=\SI{1429}{\nm}$).
In the photonic stopband, the phase is seen to decrease from $+\pi$ to $-\pi$.
To synthesize noisy data, we add normally distributed random numbers with a standard deviation $\sigma=0.002$ to \emph{both} the real and imaginary parts of the synthetic complex reflectivity amplitude.
\par %%%%%%%%%%%%%%%%%%%%%
The real and imaginary part of the complex reflectivity, as shown in figure~\ref{fig:wae_comparison}\,(a), are smoothed together.
This is important for finding the weights $w_i$ using our cross-validation metric, because the Whittaker--Henderson smoother itself is a linear operation on the data.
As a roughness measure, we use a linear combination of a fourth and second finite difference as
%%%%%%%%%%%%%%%%%%%%%%%%%%
\begin{equation}
    \mathbf{D}=\Delta^{(4)}_{\tilde{\nu}}-\frac{1}{2}\Delta^{(2)}_{\tilde{\nu}}.
    \label{eq:roughness_bragg}
\end{equation}
%%%%%%%%%%%%%%%%%%%%%%%%%%
Our choice of roughness measure is largely empirical and based on intuition, it should therefore not be considered as a universal recommendation.
A simple second or third difference roughness measure might be sufficient in case no derivative is taken.
Here, we take the second derivative of noisy data, therefore we decided to have at least a fourth difference roughness measure so that high-frequency noise is sufficiently suppressed.
We found that this tends to slightly oversmooth the data, to combat this we added a second difference to the roughness measure.
The second difference is dominant for low frequency components, compared to the fourth difference, reducing oversmoothing.
When combining roughness measures, we must take care not to exaggerate specific features, as discussed in subsection~\ref{subsec:roughness_meas}.
In case of combining a fourth and second difference, a resonance can appear in the frequency response of the smoother.
To avoid this, we require the sign of the second difference term to be negative.
Furthermore, we found that adding a factor one half to the second difference gives better results, leading to the empirically found roughness measure in equation~\ref{eq:roughness_bragg}.
\par %%%%%%%%%%%%%%%%%%%%%
Figure~\ref{fig:wae_comparison}\,(b) shows the optimal weights $w_i$ that minimize the cross-validation metric.
The smoothness parameter is fixed to $\alpha=\num{1}$.
The weights $w_i$ are divided into 80 knots, where the spline coefficient of each knot is found using a simple multivariate minimization method, as described in subsection~\ref{subsec:find_weights}.
The resulting weights $w_i$ span a range of more than four orders of magnitudes, clearly showing that a single smoothing parameter is not sufficient for smoothing the data.
As expected, we find that the weights are highest where the data behaves more erratically.
A high weight is needed for the smoothed result $\mathbf{z}$ to more closely follow the input data $\mathbf{y}$.
The difference between the smoothed result and the ground truth is shown in figure~\ref{fig:wae_comparison}\,(c).
The error is mostly smaller than the standard deviation of the added noise, showing that no oversmoothing is occurring.
The amount of high-frequency noise is strongly reduced in the smoothed data, which is important when calculating derivatives from data.
\par %%%%%%%%%%%%%%%%%%%%%
The four panels in figure~\ref{fig:method_gdd_comparison} compare four different ways to determine the group delay dispersion $D_2(\omega)$ from synthetic data with noise presented in figure~\ref{fig:wae_comparison}, and thus form a central result of this section.
The four panels in appendix figure~\ref{fig:method_residual_comparison} show the respective residuals between the calculated group delay dispersion and the ground-truth dispersion that are shown in figure~\ref{fig:method_gdd_comparison}.
Figure~\ref{fig:method_gdd_comparison}\,(a) shows the group delay dispersion that is calculated from noisy synthetic reflectivity data without smoothing.
The mean absolute error (MAE) of the dispersion without smoothing compared to the ground truth is \SI{5259}{\fs\squared}.
The large amount of noise makes it impossible to discern all but the strongest features.
The first peak at \SI{7432}{\per\cm} is visible, as is the second at \SI{7600}{\per\cm} and faintly the third at \SI{7800}{\per\cm}, but all subsequent features are obscured by the noise.
\par %%%%%%%%%%%%%%%%%%%%% 
In figure~\ref{fig:method_gdd_comparison}\,(b), the noisy reflectivity data is first filtered by a fourth-order Savitzky--Golay filter with an optimal window length of nine points, before the group delay dispersion is determined.
The filter is able to reproduce all features, except the sharpest $D_2(\omega)$ feature at \SI{7432}{\per\cm}, which is underestimated by \SI{14}{\percent}.
However, a single smoothing parameter, the kernel width, does not allow to optimally reduce noise and reproduce sharp features.
Furthermore, Savitzky--Golay filters are not good at reducing high-frequency noise~\cite{Schmid2022MSAu}, which severely hampers the effectiveness of the filter when calculating derivatives from noisy data.
\par %%%%%%%%%%%%%%%%%%%%%
To better assess the performance of the Savitzky--Golay filter, we compare the result of a third-, fourth-, fifth- and sixth-order filter.
The optimal window length of each filter is found using generalized cross-validation.
All filters give visually very similar results and therefore only the result from the fourth-order filter is plotted. 
For the third-order filter we find an optimal window length of five points, which gives an MAE of \SI{1982}{\fs\squared}.
For the fourth-order filter we find an optimal window length of nine points, which gives an MAE of \SI{1608}{\fs\squared}.
For the fifth-order filter we find an optimal window length of nine points, which gives an MAE of \SI{1612}{\fs\squared}.
For the sixth-order filter we find an optimal window length of thirteen points, which gives an MAE of \SI{1425}{\fs\squared}.
Even higher order filters barely improve the MAE metric and increase ringing and are therefore inadequate.
\par %%%%%%%%%%%%%%%%%%%%%
In figure~\ref{fig:method_gdd_comparison}\,(c), the noisy reflectivity data is first filtered by a normal Whittaker--Henderson smoother with a fourth-order roughness penalty and optimal smoothness parameter $\alpha=\num{0.110}$, before the group delay dispersion is determined.
The Whittaker--Henderson smoother is much better at suppressing high-frequency noise than the Savitzky--Golay filter and is therefore better suited when calculating derivatives from noisy data.
The resulting smoothed group delay dispersion has much less noise and more accurately follows the peaks and valleys compared to the Savitzky--Golay filter.
The filter is also able to reproduce all features, except the sharpest $D_2(\omega)$ feature at \SI{7432}{\per\cm}, which is underestimated by \SI{11}{\percent}.
\par %%%%%%%%%%%%%%%%%%%%%
To better assess the performance of the Whittaker--Henderson filter, we compare the result of a third-, fourth-, fifth- and sixth-order roughness measure.
The optimal smoothing parameter $\alpha$ of each filter is found using generalized cross-validation.
All smoothers give visually very similar results and therefore only the result from the smoother with a fourth-order roughness measure is plotted.
For the third-order roughness measure we find an optimal smoothing parameter $\alpha\approx\num{0.118}$, which gives an MAE of \SI{1569}{\fs\squared}.
For the fourth-order roughness measure we find an optimal smoothing parameter $\alpha\approx\num{0.110}$, which gives an MAE of \SI{1240}{\fs\squared}.
For the fifth-order roughness measure we find an optimal smoothing parameter $\alpha\approx\num{0.085}$, which gives an MAE of \SI{1162}{\fs\squared}.
For the sixth-order roughness measure we find an optimal smoothing parameter $\alpha\approx\num{0.061}$, which gives an MAE of \SI{1146}{\fs\squared}.
For the same order filter, we find that the Whittaker--Henderson filter has significantly better performance than the Savitzky--Golay filter.
Even higher order roughness measures barely improve the MAE metric and increase ringing and are therefore inadequate.
\par %%%%%%%%%%%%%%%%%%%%%
In figure~\ref{fig:method_gdd_comparison}\,(d), the noisy reflectivity data is first reconstructed by our weighted Whittaker--Henderson smoothing method, before the group delay dispersion is determined.
The continuously variable smoothing weights $w_i$ allow for a good balance between noise reduction and faithful reproduction of the group delay dispersion for both sharp and broad peaks and valleys.
To study the effects of the number of knots we compare the results of 41, 80 and 160 knots respectively.
The resulting optimal weights $w_i$ for each number of knots are shown in figure~\ref{fig:knots_comparison}.
The results are visually very similar and therefore only the result of the smoother with 80~knots is plotted. 
\par %%%%%%%%%%%%%%%%%%%%%
When using 41~knots we find an MAE of only \SI{278}{\fs\squared}, significantly better than the single parameter Savitzky--Golay filter and Whittaker--Henderson smoother.
The root-mean-squared error (RMSE) is \SI{869}{\fs\squared}.
All features are accurately reconstructed, however due to the limited degrees of freedom of the weights $w_i$ some undersmoothing occurs near sharp features.
The optimization process takes on average about \SI{1.2}{\s}.
When using 80~knots we find a slightly better MAE of only \SI{254}{\fs\squared} and an RMSE of \SI{893}{\fs\squared}
This is more than 20 times smaller than the MAE of the dispersion without smoothing.
The optimization process for 80~knots takes on average about \SI{1.9}{\s}.
When further increasing to 160~knots we find an MAE of \SI{242}{\fs\squared} and an RMSE of \SI{924}{\fs\squared}.
The weights $w_i$ now have too many degrees of freedom, leading to oversmoothing, as is apparent from the strongly increased RMSE.
The optimization process for 160~knots takes on average about \SI{3.7}{\s}.
\par %%%%%%%%%%%%%%%%%%%%%
Using the noise variances, we do an error analysis of the smoothed dispersion.
By calculating the covariance matrix from equation~\ref{eq:wh_covariance}, we estimate the unbiased credibility interval of the smoothed complex reflectivity $r$.
We then use error propagation to estimate the credibility interval $\sigma_{\mathbf{z}}$ of the smoothed group delay dispersion.
When using 80~knots, the estimated credibility interval $\sigma_{\mathbf{z}}$ of the smoothed dispersion for all data points is on average about \num{41} times smaller than the error of the dispersion without smoothing.
We find that \SI{68.9}{\percent} of all smoothed dispersion data points agree with the ground-truth dispersion within the $1\sigma_{\mathbf{z}}$ credibility interval.
Within the $2\sigma_{\mathbf{z}}$ credibility interval we find no less than \SI{95.8}{\percent} agreement, and for the $3\sigma_{\mathbf{z}}$ credibility interval we find a strikingly high \SI{99.8}{\percent} agreement.
All smoothed dispersion values are within $4\sigma_{\mathbf{z}}$ of the ground-truth dispersion.
The agreement percentages are close to the expected values for normally distributed errors, showing a faithful reconstruction from noisy data to a smooth curve, and thus illustrating the power of our method for our original goal. 
%%%%%%%%%%%%%%%%%%%%%%%%%%%%%%%%%%%%%%%
\section{Smoothing unequally Spaced Data}\label{sec:uneven_spaced}
%%%%%%%%%%%%%%%%%%%%%%%%%%%%%%%%%%%%%%%
\par %%%%%%%%%%%%%%%%%%%%%
An important source of difficulties in interpreting data arises when the data points are unequally spaced, as is the case for a significant part of real-world data.
Smoothing unequally spaced data often involves changing window widths over the data or resampling data points, which is messy, and prone to artifacts~\cite{Press2007book}. 
These issues do not exist for Whittaker--Henderson smoothing, but we have to take two considerations into account.
Firstly, it is vital to properly calculate the finite difference coefficients of the matrix $\mathbf{D}$ for each data point in $\mathbf{y}$.
Here it is necessary to know the (relative) location of each data point.
Secondly, in case that the feature sizes do not scale with the sampling density, we have to make sure that each part of the smooth curve $\mathbf{z}$ contributes equally to the roughness penalty. 
Therefore, we define a scaled Whittaker--Henderson smoother that is based on minimizing a functional $Q_{\rm s}[\mathbf{z}]$ defined as
%%%%%%%%%%%%%%%%%%%%%%%%%%
\begin{equation}
    Q_{\rm s}[\mathbf{z}] \equiv \sum_{i=1}^N {w_i \abs{y_i - z_i}^2\Delta_i} 
    + \alpha \sum_{i=1}^N {\abs{({\rm D}z)_i}^2\Delta_i}, 
    \label{eq:uneven_Q}
\end{equation}
%%%%%%%%%%%%%%%%%%%%%%%%%%
where $\Delta_i$ is the step size for each respective data point $i$.
This can be understood as balancing the average squared difference $\abs{y_i - z_i}^2$ against the average roughness $\abs{({\rm D}z)_i}^2$.
The smooth result $\mathbf{z}$ minimizing this functional is found by solving
%%%%%%%%%%%%%%%%%%%%%%%%%%
\begin{equation}
    \left(\mathbf{SW} + \alpha\mathbf{D}^{\rm T} \mathbf{SD}\right)\mathbf{z} = \mathbf{SW}\mathbf{y}, 
    \label{eq:wh_scaled}
\end{equation}
%%%%%%%%%%%%%%%%%%%%%%%%%%
with $\mathbf{S}$ a diagonal scaling matrix consisting of the step sizes $\Delta_i$ for all data points, 
%%%%%%%%%%%%%%%%%%%%%%%%%%
\begin{equation}
    \mathbf{S} \equiv \text{diag}\{\Delta_1, \Delta_2,\dots,\Delta_N\}, 
    \label{eq:S_matrix}
\end{equation}
%%%%%%%%%%%%%%%%%%%%%%%%%%
and $\mathbf{W}$ the diagonal weight matrix as defined in equation~\ref{eq:wh_weights}.
In case all step sizes $\Delta_i$ are equal, \textit{i.e.}, for equally spaced data, this reduces back to weighted Whittaker--Henderson smoothing as defined in equation~\ref{eq:wh_weighted_linear_system}.
This linear system consists of sparse matrices, like normal Whittaker--Henderson smoothing, and is thus also very efficient to solve.
The scaled Whittaker--Henderson smoother can also be used to smoothly interpolate the unequally spaced data onto an equally spaced grid using the weights $w_i$.
%%%%%%%%%%%%%%%%%%%%%%%%%%
\begin{figure}
    \centering
    \includegraphics[width=\linewidth]{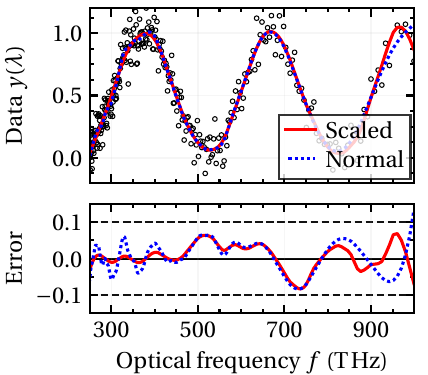}
    \caption{
        Example of smoothing unequally spaced noisy data.
        The data is equally spaced as a function of wavelength $\lambda$ and plotted as a function of the optical frequency $f$.
        The red line is the result of scaled Whittaker--Henderson smoothing in the optical frequency domain.
        The blue dotted line is the result of normal Whittaker--Henderson smoothing in the wavelength domain.
        The standard deviation (dashed lines) of the added noise is $\sigma=0.1$.
        }
    \label{fig:uneven_spaced}
\end{figure}
%%%%%%%%%%%%%%%%%%%%%%%%%%
%%%%%%%%%%%%%%%%%%%%%%%%%%
\subsection{Mapped Unequally Spaced Data}\label{subsec:application_smooth_nu_sampling}
%%%%%%%%%%%%%%%%%%%%%%%%%%
\par %%%%%%%%%%%%%%%%%%%%%
The first type of unequally spaced data we consider is the situation where the data is equally spaced in one $x_{\rm e}$-domain, but where we want to know the features as a function of the unequally spaced $x_{\rm u}$-domain.
An example is data that are equally sampled in wavelength whereas we want to obtain information in the optical frequency domain.
When the mapping from $x_{\rm e}$ to $x_{\rm u}$ is given by the known and smooth function $m$ as
%%%%%%%%%%%%%%%%%%%%%%%%%%
\begin{equation}
    x_{\rm u}=m(x_{\rm e}), 
    \label{eq:mapping_xs_xm}
\end{equation}
%%%%%%%%%%%%%%%%%%%%%%%%%%
then it is possible to approximate the scaling matrix $\mathbf{S}$ as
%%%%%%%%%%%%%%%%%%%%%%%%%%
\begin{equation}
    \mathbf{S} = \text{diag}\{m'(x_{{\rm e},1}), m'(x_{{\rm e},2}), \dots, m'(x_{{\rm e},N})\}. 
    \label{eq:S_matrix_smooth}
\end{equation}
%%%%%%%%%%%%%%%%%%%%%%%%%%
The roughness matrix $\mathbf{D}$ has to be calculated using a roughness penalty in the unequally spaced $x_{\rm u}$-domain.
For example, a second-order roughness penalty in $x_{\rm u}$-domain is expressed as
%%%%%%%%%%%%%%%%%%%%%%%%%%
\begin{equation}
    \frac{\dd[2]{z(x_{\rm e})}}{\dd{x_{\rm u}^2}}=\frac{1}{m'(x_{\rm e})^2}\frac{\dd[2]{z(x_{\rm e})}}{\dd{x_{\rm e}^2}} - \frac{m''(x_{\rm e})}{m'(x_{\rm e})^3}\frac{\dd{z(x_{\rm e})}}{\dd{x_{\rm e}}},
    \label{eq:mapping_dy}
\end{equation}
%%%%%%%%%%%%%%%%%%%%%%%%%%
using derivatives in the equally spaced $x_{\rm e}$-domain.
\par %%%%%%%%%%%%%%%%%%%%%
An example of a real-world application is when we use a spectrometer that measures an optical spectrum in units of nanometers, but we want to know the spectrum in optical frequency units.
We then have a mapping from wavelength $\lambda$ to optical frequency $f$ given by
%%%%%%%%%%%%%%%%%%%%%%%%%%
\begin{equation}
    f(\lambda) = c/\lambda. 
    \label{eq:mapping_freq_to_wl}
\end{equation}
%%%%%%%%%%%%%%%%%%%%%%%%%%
The roughness penalty from equation~\ref{eq:mapping_dy} becomes
%%%%%%%%%%%%%%%%%%%%%%%%%%
\begin{equation}
    \frac{\dd[2]{z(\lambda)}}{\dd{f^2}}=\frac{\lambda^4}{c^2}\frac{\dd[2]{z(\lambda)}}{\dd{\lambda^2}} 
    + \frac{\lambda^3}{c^2}\frac{\dd{z(\lambda)}}{\dd{\lambda}}, 
    \label{eq:mapping_df}
\end{equation}
%%%%%%%%%%%%%%%%%%%%%%%%%%
which is approximated using finite differences.
\par %%%%%%%%%%%%%%%%%%%%%
An example is shown in figure~\ref{fig:uneven_spaced}, where the synthetic data is equally spaced as a function of wavelength, but plotted as a function of optical frequency.
The synthetic data is a sine wave with a period of \SI{300}{\tera\hertz}, sampled from wavelength $\lambda=\SIrange{300}{1200}{\nm}$ at an interval of $\delta\lambda=\SI{3}{\nm}$.
To simulate noisy data, we add normally distributed random numbers with a standard deviation of $\sigma=0.1$ to the ground-truth data.
Both smoothers use a second finite difference as roughness measure, in wavelength units and optical frequency units respectively.
When smoothing in wavelength units, using the normal Whittaker--Henderson smoother, we find an optimal smoothing parameter $\alpha_{\lambda}\approx\num{1.47e4}$ and an MAE of \num{0.030}.
When smoothing in optical frequency units, using the scaled Whittaker--Henderson smoother, we find an optimal smoothing parameter $\alpha_{f}\approx\num{4.24e4}$ and a much smaller MAE of \num{0.019}.
The result from the scaled smoother is equally smooth over the entire optical frequency axis, while the result from the normal smoother shows both over- and undersmoothing.
Oversmoothing occurs on the right, sparsely sampled, side of the data and causes the smoothed result to no longer accurately follow the trends of the data.
Undersmoothing occurs on the left, densely sampled, side of the data and causes the smoothed result to fluctuate too much due to insufficient noise suppression.
%%%%%%%%%%%%%%%%%%%%%%%%%%
\begin{figure}[t]
    \centering
    \includegraphics[width=\linewidth]{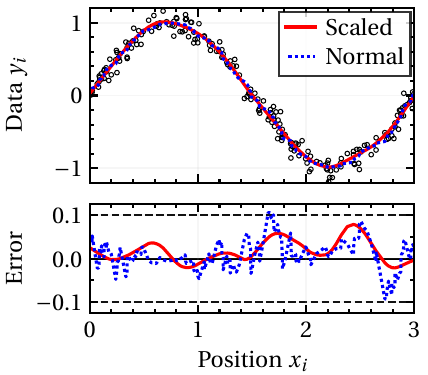}
    \caption{
        Example of smoothing randomly spaced noisy data $y_i$ sampled at positions $x_i$.
        The red line is the result of scaled Whittaker--Henderson smoothing with properly calculated derivatives.
        The blue dotted line is the result of normal Whittaker--Henderson smoothing that assumes equally spaced data.
        The standard deviation (dashed lines) of the added noise is $\sigma=0.1$.
        }
    \label{fig:random_spaced}
\end{figure}
%%%%%%%%%%%%%%%%%%%%%%%%%%
%%%%%%%%%%%%%%%%%%%%%%%%%%
\subsection{Randomly Spaced Data}\label{subsec:application_random_sampling}
%%%%%%%%%%%%%%%%%%%%%%%%%%
\par %%%%%%%%%%%%%%%%%%%%%
A second type of unequally spaced data we consider is where the data is sampled at random positions in the $x$-domain.
As a consequence, we have to estimate the finite difference coefficients at each position $x_i$ separately.
An approximate form of a first finite difference for arbitrarily spaced data points is
%%%%%%%%%%%%%%%%%%%%%%%%%%
\begin{equation}
    z_i'\approx \frac{z_{i}-z_{i-1}}{x_{i}-x_{i-1}}, 
    \label{eq:random_spaced_1st_difference}
\end{equation}
%%%%%%%%%%%%%%%%%%%%%%%%%%
and an approximate second difference form is
%%%%%%%%%%%%%%%%%%%%%%%%%%
\begin{equation}
        z_i''\approx \frac{2}{x_{i+1}-x_{i-1}} \left[\frac{z_{i+1}-z_{i}}{x_{i+1}-x_{i}}-\frac{z_{i}-z_{i-1}}{x_{i}-x_{i-1}}\right]. 
    \label{eq:random_spaced_2nd_difference}
\end{equation}
%%%%%%%%%%%%%%%%%%%%%%%%%%
\par %%%%%%%%%%%%%%%%%%%%%
An example is shown in figure~\ref{fig:random_spaced}.
The synthetic data is a single period of a sine wave, on an interval from $x=\num{0}\text{ to }\num{3}$, sampled at 200 random positions.
To simulate noisy data we add normally distributed random numbers with a standard deviation of $\sigma=0.1$ to the ground-truth data.
For the normal Whittaker--Henderson smoother we find an optimal smoothing parameter $\alpha\approx\num{5.10e-5}$ and an MAE of \num{0.027}.
For the scaled Whittaker--Henderson smoother we find an optimal smoothing parameter $\alpha\approx\num{8.47e-5}$ and a slightly smaller MAE of \num{0.023}.
Both smoothers use a second finite difference as roughness measure.
The result from the scaled smoother is equally smooth over the entire $x$-axis, while the result from the normal smoother shows many sudden steps.
These sudden steps are due to data points being very close together, causing a local underestimation in the roughness metric for normal Whittaker--Henderson smoothing.
This simple fact makes it impossible for the normal smoother to make a smooth curve from randomly sampled data.
Care has to be taken when two data points are extremely close together compared to other data points, as this can lead to numerical instabilities.
%%%%%%%%%%%%%%%%%%%%%%%%%%%%%%%%%%%%%%%
\section{Smoothing Discontinuous data}\label{sec:discontinuous}
%%%%%%%%%%%%%%%%%%%%%%%%%%%%%%%%%%%%%%%
\par %%%%%%%%%%%%%%%%%%%%%
A major challenge in data analysis occurs when discontinuities exist, either in the original data $\mathbf{y}$ or in its derivatives.
Discontinuities are associated with strong high-frequency components, which makes them exceedingly difficult to properly smooth.
Simply splitting the data in parts at a discontinuity does not take advantage of information on both sides of the discontinuity and leads to large errors.
In this section we distinguish two important types of discontinuities, but many more kinds can be implemented using our smoother.
%%%%%%%%%%%%%%%%%%%%%%%%%%
\begin{figure}
    \centering
    \includegraphics[width=\linewidth]{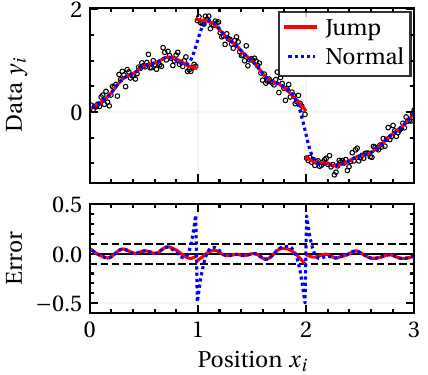}
    \caption{
        Example of smoothing noisy data $y_i$ sampled at positions $x_i$, with two jump discontinuities.
        The red line is the result of Whittaker--Henderson smoothing without penalty for a jump at the discontinuities.
        The blue dotted line is the result of normal Whittaker--Henderson smoothing.
        The standard deviation (dashed lines) of the added noise is $\sigma=0.1$.
    }
    \label{fig:jump_smoothing}
\end{figure}
%%%%%%%%%%%%%%%%%%%%%%%%%%
%%%%%%%%%%%%%%%%%%%%%%%%%%
\subsection{Continuous First Derivative: Jumps}\label{subsec:application_jumps}
%%%%%%%%%%%%%%%%%%%%%%%%%%
\par %%%%%%%%%%%%%%%%%%%%%
A first type of discontinuity is when there is a jump in the data at some point $x_{\rm d}$, whereas the derivative is continuous, \textit{i.e.}, when
%%%%%%%%%%%%%%%%%%%%%%%%%%
\begin{equation}
    y(x_{\rm d}^-)\neq y(x_{\rm d}^+)
    \label{eq:define_jump1}
\end{equation}
%%%%%%%%%%%%%%%%%%%%%%%%%%
and
%%%%%%%%%%%%%%%%%%%%%%%%%%
\begin{equation}
    y'(x_{\rm d}^-)= y'(x_{\rm d}^+).
    \label{eq:define_jump2}
\end{equation}
%%%%%%%%%%%%%%%%%%%%%%%%%%
The information on the existence of a jump at $x_{\rm d}$ can be included in the roughness matrix $\mathbf{D}$.
An example of a second finite difference matrix with jump discontinuity is
%%%%%%%%%%%%%%%%%%%%%%%%%%
\begin{equation}
    \mathbf{D} = \frac{1}{\Delta^2}
    \left[ {\begin{array}{rrrr|rrrr}
        1 & -2 &               1 &              0 &              0 &               0 &  0 &      0 \\
        0 &  1 &              -2 &         \phm 1 &              0 &               0 &  0 &      0 \\
        0 &  0 & \y -\frac{1}{2} & \y \frac{1}{2} & \y \frac{1}{2} & \y -\frac{1}{2} &  0 &      0 \\
          &    &                 &                &                & \vspace{-.36cm} &    &        \\
        0 &  0 & \y -\frac{1}{2} & \y \frac{1}{2} & \y \frac{1}{2} & \y -\frac{1}{2} &  0 &      0 \\
        0 &  0 &               0 &              0 &         \phm 1 &              -2 &  1 &      0 \\
        0 &  0 &               0 &              0 &              0 &               1 & -2 & \phm 1 \\
    \end{array} } \right]. 
    \label{eq:D_mat_jump}
\end{equation}
%%%%%%%%%%%%%%%%%%%%%%%%%%
At the yellow highlighted cells, the second finite difference is replaced by a different kind of roughness measure.
This roughness measure has a penalty when the slopes $\mathbf{y}'$ at both sides of the discontinuity differ, yet there is no penalty when the values $\mathbf{y}$ differ at both sides of the discontinuity.
\par %%%%%%%%%%%%%%%%%%%%%
A comparison of two equivalent Whittaker--Henderson smoothers, one with jump discontinuities and one without, is shown in figure~\ref{fig:jump_smoothing}.
The synthetic data is a single period of a sine wave, on an interval from $x=\num{0}\text{ to }\num{3}$, sampled on a uniform grid of 200 data points.
At the index $x=1$ the data jumps up by one, at the index $x=2$ the data jumps back down by one.
To simulate noisy data, we add normally distributed random numbers with a standard deviation of $\sigma=0.1$ to the ground-truth data.
For the Whittaker--Henderson smoother with jump discontinuity we find an optimal smoothing parameter $\alpha\approx\num{3.86e-6}$.
For the normal Whittaker--Henderson smoother we use the same smoothing parameter, to compare the effects of adding the jump discontinuity.
Both smoothers use a second finite difference as roughness measure.
The smoother with discontinuities faithfully reproduces the true jumps in the curve.
The error stays within the standard deviation of the added noise and is independent of the size of the jump.
The smoother without discontinuity removes all high frequencies from the jumps, which turns the jumps into smooth slopes, and thus loses the essential features, namely the discontinuities.
The error is much larger than the standard deviation of the added noise, and scales with the size of the jump.
%%%%%%%%%%%%%%%%%%%%%%%%%%
\begin{figure}
    \centering
    \includegraphics[width=\linewidth]{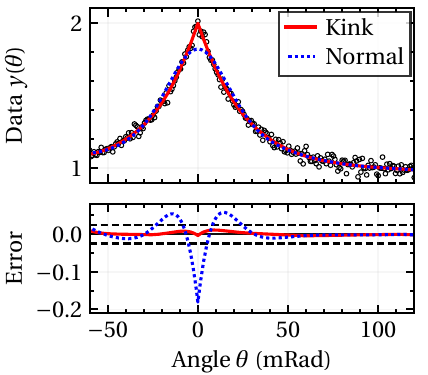}
    \caption{
        Example of smoothing noisy, synthetic, "enhanced backscattering cone" data with a kink discontinuity.
        The light intensity $y$ is plotted as a function of the backscatter angle $\theta$.
        The red line is the result of Whittaker--Henderson smoothing without penalty for a kink at the discontinuity.
        The blue dotted line is the result of normal Whittaker--Henderson smoothing.
        The standard deviation (dashed lines) of the added noise is $\sigma=\num{0.025}$.
    }
    \label{fig:kink_smoothing}
\end{figure}
%%%%%%%%%%%%%%%%%%%%%%%%%%
%%%%%%%%%%%%%%%%%%%%%%%%%%
\subsection{Discontinuous First Derivative: Enhanced Backscattering Cone}\label{subsec:application_ebsc}
%%%%%%%%%%%%%%%%%%%%%%%%%%
\par %%%%%%%%%%%%%%%%%%%%%
A second type of discontinuity is when the first derivative is discontinuous at some point $x_{\rm d}$, whereas the curve itself is continuous, \textit{i.e.}, when
%%%%%%%%%%%%%%%%%%%%%%%%%%
\begin{equation}
    y(x_{\rm d}^-) = y(x_{\rm d}^+)
    \label{eq:define_kink1}
\end{equation}
%%%%%%%%%%%%%%%%%%%%%%%%%%
and
%%%%%%%%%%%%%%%%%%%%%%%%%%
\begin{equation}
    y'(x_{\rm d}^-) \neq y'(x_{\rm d}^+).
    \label{eq:define_kink2}
\end{equation}
%%%%%%%%%%%%%%%%%%%%%%%%%%
This situation is also known as a kink and a practical situation where this occurs is at the center of an enhanced backscattering cone~\cite{Wiersma1995RSI}.
Enhanced backscattering is a remarkable interference phenomenon that occurs in multiple scattering of light and other waves, where counterpropagating paths constructively interfere, leading to an intensity enhancement into the exact backscattering direction up to a factor two, compared to the diffuse background intensity.
Moving away from the exact backscattering direction, the phases between the counterpropagating paths decohere, leading to a gradual decrease in enhancement, hence why it is named a cone.
In the absence of absorption, the backscattering cone decays approximately exponentially as a function of the backscattering angle.
The cone is then not analytic at the exact backscattering angle but is a cusp.
\par %%%%%%%%%%%%%%%%%%%%%
The information of the existence of a kink at $x_{\rm d}$ can be included in the roughness matrix $\mathbf{D}$.
An example of a second finite difference matrix with kink discontinuity is
%%%%%%%%%%%%%%%%%%%%%%%%%%
\begin{equation}
    \mathbf{D} = \frac{1}{\Delta^2}
    \left[ {\begin{array}{rrr|r|rrr}
        1 & -2 &     1 &      0 &     0 &  0 &       0 \\
        0 &  1 &    -2 &      1 &     0 &  0 &       0 \\
        0 &  0 &  \y 0 &   \y 0 &  \y 0 &  0 &       0 \\
        0 &  0 &     0 & \phm 1 &    -2 &  1 &       0 \\
        0 &  0 &     0 &      0 &     1 & -2 &  \phm 1 \\
    \end{array} } \right].
    \label{eq:D_mat_kink}
\end{equation}
%%%%%%%%%%%%%%%%%%%%%%%%%%
The yellow highlighted cells are set to zero, instead of to the second finite difference coefficients.
This means that there is no penalty for smoothness at the position of the discontinuity, and a kink is free to appear in $\mathbf{z}$.
Since the curve surrounding the kink has to be smooth, a jump in $\mathbf{z}$ is not allowed to appear without a penalty.
\par %%%%%%%%%%%%%%%%%%%%%
A comparison of two equivalent Whittaker--Henderson smoothers, one with a kink discontinuity and one without kink, are shown in figure~\ref{fig:kink_smoothing}.
The synthetic data represents "enhanced backscattering cone" data~\cite{Albada1985PRL, Wolf1985PRL} and is a symmetrically decaying exponential with background.
It is sampled on an interval from $\theta=\SIrange{-60}{120}{\milli\radian}$ on a uniform grid of 226 data points.
At the angle $\theta=\SI{0}{\milli\radian}$ a cusp is present.
To simulate noisy data, we add normally distributed random numbers with a standard deviation of $\sigma=\SI{0.025}{\milli\radian}$ to the ground-truth data.
For the Whittaker--Henderson smoother with kink discontinuity we find an optimal smoothing parameter $\alpha\approx\num{1.54e4}$.
For the normal Whittaker--Henderson smoother we use the same smoothing parameter, to compare the effects of adding the kink discontinuity.
Both smoothers use a third finite difference as roughness measure.
The smoother with kink discontinuity faithfully reproduces the noiseless true data, including the cusp.
The normal Whittaker--Henderson smoother simply removes all high-frequency components, regardless of the cusp, and is thus unable to reproduce the noiseless true data.
%%%%%%%%%%%%%%%%%%%%%%%%%%%%%%%%%%%%%%%
\section{Smoothing with Boundary Conditions}\label{sec:boundary_cond}
%%%%%%%%%%%%%%%%%%%%%%%%%%%%%%%%%%%%%%%
\par %%%%%%%%%%%%%%%%%%%%%
An omnipresent yet often neglected challenge is the smoothing of a data set at the edges of the domain.
The abrupt absence of data outside the domain often leads to large artifacts.
Boundary conditions can be naturally included in Whittaker--Henderson smoothers, leading to significantly reduced errors.
%%%%%%%%%%%%%%%%%%%%%%%%%%
\begin{figure}
    \centering
    \includegraphics[width=\linewidth]{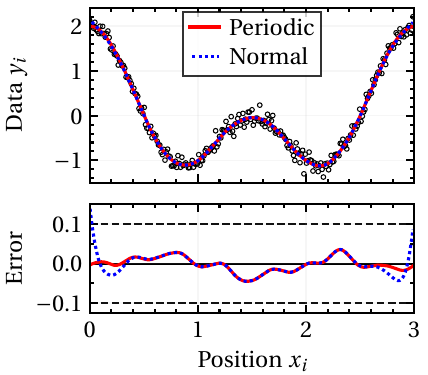}
    \caption{
        Example of smoothing noisy periodic data $y_i$ sampled at positions $x_i$.
        The red line is the result of a Whittaker--Henderson smoother with periodic boundary conditions.
        The blue dotted line is the result of a normal Whittaker--Henderson smoother.
        The standard deviation of the added noise is $\sigma=0.1$.
    }
    \label{fig:periodic_bc}
\end{figure}
%%%%%%%%%%%%%%%%%%%%%%%%%%
%%%%%%%%%%%%%%%%%%%%%%%%%%
\subsection{Periodic Boundary Conditions}\label{subsec:periodic_boundary}
%%%%%%%%%%%%%%%%%%%%%%%%%%
\par %%%%%%%%%%%%%%%%%%%%%
One of the simplest kinds of boundary conditions is a periodic boundary condition.
This poses no issues for normal convolution-based filters, as the data is simply repeated.
For Whittaker--Henderson smoothers it is only a matter of defining a roughness penalty that connects one side of the data to the other side.
An example of a second finite difference matrix with a periodic boundary condition is
%%%%%%%%%%%%%%%%%%%%%%%%%%
\begin{equation}
    \mathbf{D} = \frac{1}{\Delta^2}
    \left[ {\begin{array}{rrrrr}
        -2 &  1 &  0 &  0 & \y 1\\
         1 & -2 &  1 &  0 &    0\\
         0 &  1 & -2 &  1 &    0\\
         0 &  0 &  1 & -2 &    1\\
      \y 1 &  0 &  0 &  1 &   -2\\
    \end{array} } \right], 
    \label{eq:D_periodic}
\end{equation}
%%%%%%%%%%%%%%%%%%%%%%%%%%
where $\Delta$ is the finite difference step size.
The yellow highlighted cells show where the finite difference coefficients wrap around from one side of the $\mathbf{D}$ matrix to the other, which enforces the periodicity.
\par %%%%%%%%%%%%%%%%%%%%%
An example of smoothing with periodic boundary conditions is shown in figure~\ref{fig:periodic_bc}.
We compare a Whittaker--Henderson smoother without boundary conditions to an otherwise equivalent Whittaker--Henderson smoother with periodic boundary conditions. 
Both smoothers use a second finite difference, similar to equation~\ref{eq:D_periodic}, as roughness measure.
For the Whittaker--Henderson smoother with periodic boundary conditions we find an optimal smoothing parameter $\alpha\approx\num{3.34e-5}$.
For the normal Whittaker--Henderson smoother we use the same smoothing parameter, to compare the effects of adding the boundary condition.
The errors of the smoother with periodic boundary conditions are about the same order of magnitude everywhere, and it does not suffer from boundary artifacts.
The smoother without boundary conditions, on the other hand, has significantly larger errors, and is also not continuous at the boundaries.
%%%%%%%%%%%%%%%%%%%%%%%%%%
\subsection{Dirichlet Boundary Conditions}\label{subsec:dirichlet_boundary}
%%%%%%%%%%%%%%%%%%%%%%%%%%
\par %%%%%%%%%%%%%%%%%%%%%
When a precise value at a boundary is known, we enforce this condition using a Dirichlet boundary condition.
Usual convolution filters apply this boundary condition by padding the edges of the data with chunks of the known value.
This is not necessary for (weighted) Whittaker--Henderson smoothers.
As an example, we consider the case of a weighted Whittaker--Henderson smoother where it is known that $z_1=a$.
We then have to solve a linear system defined as
%%%%%%%%%%%%%%%%%%%%%%%%%%
\begin{equation}
    \left[ {\begin{array}{ccccc}
        1      & 0      & 0      & \dots\\
        c_{21} & c_{22} & c_{23} & \dots\\
        c_{31} & c_{32} & c_{33} & \dots\\
        c_{41} & c_{42} & c_{43} & \dots\\
        \vdots & \vdots & \vdots & \ddots\\
    \end{array} } \right]
    \left[ {\begin{array}{c}
        z_1\\ z_2 \\ z_3\\ z_4\\ \vdots\\
    \end{array} } \right] =   
    \left[ {\begin{array}{c}
        a\\ w_2 y_2 \\ w_3 y_3\\ w_4 y_4\\ \vdots\\
    \end{array} } \right], 
    \label{eq:matrix_dirichlet}
\end{equation}
%%%%%%%%%%%%%%%%%%%%%%%%%%
where the $c_{ij}$ elements are given by the $\mathbf{C}$ matrix, defined as
%%%%%%%%%%%%%%%%%%%%%%%%%%
\begin{equation}
    \mathbf{C} = \mathbf{W} + \alpha \mathbf{D}^{\rm T} \mathbf{D}. 
    \label{eq:define_C}
\end{equation}
%%%%%%%%%%%%%%%%%%%%%%%%%%
This means that this boundary condition only requires us to change the first row of the matrix $\mathbf{C}$ and vector $\mathbf{Wy}$ and is thus just as efficient to solve as the weighted Whittaker--Henderson smoother without boundary conditions.
Solving this system is equivalent to finding the smooth curve $\mathbf{z}$ minimizing the functional $Q[\mathbf{z}]$, while enforcing that $z_1=a$.
%%%%%%%%%%%%%%%%%%%%%%%%%%
\subsection{Neumann Boundary Conditions}\label{subsec:neumann_boundary}
%%%%%%%%%%%%%%%%%%%%%%%%%%
\par %%%%%%%%%%%%%%%%%%%%%
When the slope of the data at a boundary is known, we enforce this condition using a Neumann boundary condition.
This boundary condition is straightforward to implement for Whittaker--Henderson smoothers.
As an example, we consider the case of a weighted Whittaker--Henderson smoother where it is known that $z'_1=b$.
The derivative at the left boundary is approximated using a second-order forward finite difference.
We then have to solve a linear system defined as
%%%%%%%%%%%%%%%%%%%%%%%%%%
\begin{equation}
    \left[ {\begin{array}{ccccc}
        \frac{-3}{2\Delta} & \frac{2}{\Delta} & \frac{-1}{2\Delta} & 0 & \dots\\
        c_{21} & c_{22} & c_{23} & c_{24} & \dots\\
        c_{31} & c_{32} & c_{33} & c_{34} & \dots\\
        c_{41} & c_{42} & c_{43} & c_{44} & \dots\\
        \vdots & \vdots & \vdots &\vdots & \ddots\\
    \end{array} } \right]
    \left[ {\begin{array}{c}
        z_1\\ z_2 \\ z_3\\ z_4\\ \vdots\\
    \end{array} } \right] =
    \left[ {\begin{array}{c}
        a\\ w_2 y_2 \\ w_3 y_3\\ w_4 y_4\\ \vdots\\
    \end{array} } \right],
    \label{eq:matrix_neumann}
\end{equation}
%%%%%%%%%%%%%%%%%%%%%%%%%%
where the $c_{ij}$ elements are given by the $\mathbf{C}$ matrix as defined in equation~\ref{eq:define_C} and $\Delta$ is the finite difference step size.
This has the disadvantage that we lose the data point $y_1$, yet to solve this we simply add temporary data points $y_0$ and $z_0$ which are removed again after solving the linear system.
%%%%%%%%%%%%%%%%%%%%%%%%%%
\begin{figure}
    \centering
    \includegraphics[width=\linewidth]{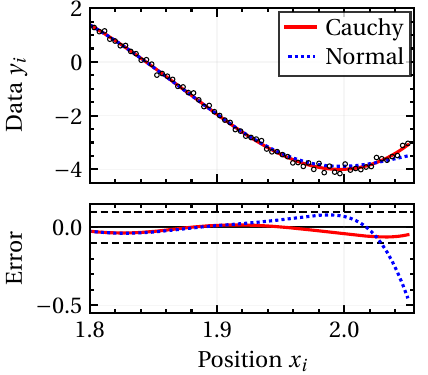}
    \caption{
        Example of smoothing noisy data $y_i$ sampled at positions $x_i$, with a difficult data boundary.
        The red line is the result of Whittaker--Henderson smoothing with Cauchy boundary conditions.
        The blue dotted line is the result of normal Whittaker--Henderson smoothing.
        The standard deviation (dashed lines) of the added noise is $\sigma=0.1$.
        }
    \label{fig:cauchy_example}
\end{figure}
%%%%%%%%%%%%%%%%%%%%%%%%%%
%%%%%%%%%%%%%%%%%%%%%%%%%%
\subsection{Cauchy Boundary Conditions}\label{subsec:cauchy_boundary}
%%%%%%%%%%%%%%%%%%%%%%%%%%
\par %%%%%%%%%%%%%%%%%%%%%
When both the precise value and the slope of the data are known at a boundary, we enforce this condition using a Cauchy boundary condition.
Even if these values are not known beforehand, we can still find an approximation by fitting a polynomial through a number of data points at the boundary.
This hybrid approach makes smoothing at the edges of data sets much more robust and reduces artifacts. 
As an example, we consider the case of a weighted Whittaker--Henderson smoother where it is known that $z_1=a$ and $z'_1=b$.
The derivative at the left boundary is approximated using a first-order forward finite difference.
We then have to solve a linear system defined as
%%%%%%%%%%%%%%%%%%%%%%%%%%
\begin{equation}
    \left[ {\begin{array}{ccccc}
        1 & 0 & 0 & \dots\\
        \frac{1}{\Delta} & \frac{-1}{\Delta} & 0 & \dots\\
        c_{31} & c_{32} & c_{33} & \dots \\
        c_{41} & c_{42} & c_{43} & \dots \\
        \vdots & \vdots & \vdots & \ddots\\
    \end{array} } \right]
    \left[ {\begin{array}{c}
        z_1\\ z_2 \\ z_3\\ z_4\\ \vdots\\
    \end{array} } \right] = 
    \left[ {\begin{array}{c}
        a\\ b \\ w_3 y_3\\ w_4 y_4\\ \vdots\\
    \end{array} } \right],
    \label{eq:matrix_cauchy}
\end{equation}
%%%%%%%%%%%%%%%%%%%%%%%%%%
where the $c_{ij}$ elements are given by the $\mathbf{C}$ matrix as defined in equation~\ref{eq:define_C} and $\Delta$ is the finite difference step size.
\par %%%%%%%%%%%%%%%%%%%%%
An example of smoothing with Cauchy boundary conditions is shown in figure~\ref{fig:cauchy_example}.
Here we compare a Whittaker--Henderson smoother with Cauchy boundary conditions to an otherwise equivalent Whittaker--Henderson smoother without boundary conditions.
The synthetic data is a chirped sine wave, on an interval from $x=\num{0}\text{ to }\num{2.05}$, sampled on a uniform grid of 500 data points.
At the right edge of the data the downward trend abruptly changes in an upward trend, which usually causes issues with smoothing.
To simulate noisy data, we add normally distributed random numbers with a standard deviation of $\sigma=0.1$ to the ground-truth data.
For the Whittaker--Henderson smoother with Cauchy boundary conditions we find an optimal smoothing parameter $\alpha\approx\num{3.86e-6}$.
For the normal Whittaker--Henderson smoother we use the same smoothing parameter, to compare the effects of adding the boundary condition.
Both smoothers use a second finite difference as roughness measure.
The Whittaker--Henderson smoother without boundary conditions suffers from boundary artifacts and is not able to accurately follow the data.
For the Whittaker--Henderson smoother with Cauchy boundary conditions we find the right boundary value and slope by fitting a second-order polynomial through the last 25 data points.
This forces the smoothed curve to follow the data more closely at the boundaries.
The error at the boundaries is then given by the error of the fit, as opposed to the error of the smoother.
This leads to significantly smaller errors, as shown by the error in figure~\ref{fig:cauchy_example}.
%%%%%%%%%%%%%%%%%%%%%%%%%%
\begin{figure*}[t]
    \centering
    \includegraphics[width=\linewidth]{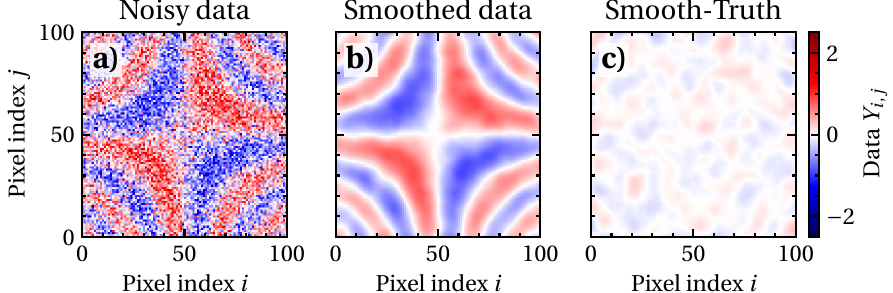}
    \caption{
        Example of multi-dimensional Whittaker--Henderson smoothing.
        \textbf{(a)} Normally distributed random numbers ($\sigma=0.5$) are added to synthetic data to simulate noise.
        \textbf{(b)} Result after smoothing.
        The amount of noise is strongly reduced, and the real trend is now clearly visible.
        \textbf{(c)} Difference between the smoothed result and the ground truth.
        The standard deviation of the difference is only $\sigma\approx\num{0.078}$, much smaller than the standard deviation of the added noise.
    }
    \label{fig:2d_whittaker_example}
\end{figure*}
%%%%%%%%%%%%%%%%%%%%%%%%%%
%%%%%%%%%%%%%%%%%%%%%%%%%%%%%%%%%%%%%%% 
\section{Multi-dimensional Smoothing}\label{sec:mult_dim_filtering}
%%%%%%%%%%%%%%%%%%%%%%%%%%%%%%%%%%%%%%%
\par %%%%%%%%%%%%%%%%%%%%%
Whittaker--Henderson smoothing is readily extended to multi-dimensional data~\cite{Moulart2014MST}, such as for instance multidimensional spectral data collected with an array detector (CCD, CMOS sensor). 
We have to flatten the multi-dimensional data into a 1D~array and calculate a roughness matrix $\mathbf{D}$ that encodes the multi-dimensional roughness measure.
For simplicity we consider two-dimensional (2D) data $Y_{i,j}$ sampled on a rectangular grid of $N_{y}$ by $N_{x}$ points respectively.
As a roughness measure, we take the Laplace operator, which is approximated using finite differences as
%%%%%%%%%%%%%%%%%%%%%%%%%%
\begin{equation}
    \begin{split}
        (\nabla^2Y)_{i,j} \approx \frac{Y_{i,j+1} -2Y_{i,j}+Y_{i,j-1}}{\Delta_x^2} \\
        + \frac{Y_{i+1,j} -2Y_{i,j}+Y_{i-1,j}}{\Delta_y^2},
    \end{split}
    \label{eq:laplace_fd}
\end{equation}
%%%%%%%%%%%%%%%%%%%%%%%%%%
where $\Delta_x$ and $\Delta_y$ are the finite difference step sizes in the $x$ and $y$ direction respectively.
The data is flattened into a one-dimensional (1D) array according to
%%%%%%%%%%%%%%%%%%%%%%%%%%
\begin{equation}
    y_{(i N_x+j)} \equiv Y_{i,j}.
    \label{eq:flat_Y}
\end{equation}
%%%%%%%%%%%%%%%%%%%%%%%%%%
Next, we calculate the elements of the roughness matrix $\mathbf{D}$ such that $(\mathbf{Dy})_{(i N_x+j)}$ is equivalent to the Laplace operator on $Y_{i,j}$ as defined in equation~\ref{eq:laplace_fd}.
By transforming the 2D problem into a 1D problem, we simply have to solve a linear system of the exact same form as the normal Whittaker--Henderson smoother defined in equation~\ref{eq:wh_linear_system}.
Since this involves sparse matrices, this is still relatively efficient to solve.
Weighted or scaled Whittaker--Henderson smoothing can also be used, depending on the problem at hand, see the previous sections.
Finally, we transform the flattened 1D data back to the 2D data, according to equation~\ref{eq:flat_Y}.
\par %%%%%%%%%%%%%%%%%%%%%
An example of 2D smoothing is shown in figure~\ref{fig:2d_whittaker_example}.
We have a grid of a 101 by 101 data points with synthetic data to which we added normally distributed noise with a standard deviation $\sigma=\num{0.5}$.
The function describing the synthetic data is
%%%%%%%%%%%%%%%%%%%%%%%%%%
\begin{equation}
    Y_{i,j} = \frac{\sin{\left(15\,(\frac{i}{50}-\frac{1}{2})(\frac{j}{50}-\frac{1}{2}) \right)}}{1 + (\frac{i}{50}-\frac{1}{2})^2 + (\frac{j}{50}-\frac{1}{2})^2}.
    \label{eq:synth_data_2d}
\end{equation}
%%%%%%%%%%%%%%%%%%%%%%%%%%
The noisy data in figure~\ref{fig:2d_whittaker_example}\,(a) is then smoothed by a normal Whittaker--Henderson smoother with the Laplacian operator as roughness measure, as defined in equation~\ref{eq:laplace_fd}.
Using generalized cross-validation we find an optimal smoothing parameter $\alpha\approx\num{18.4}$ that best balances the features against the noise.
Finally, we calculate the difference between the smoothed data and the true data without noise, which is shown in figure~\ref{fig:2d_whittaker_example}\,(c).
The standard deviation of the difference is $\sigma\approx\num{0.078}$, which is more than six times smaller than the standard deviation of the added noise, therefore we accurately resolve the underlying true data from the noise.
Multi-dimensional Whittaker--Henderson smoothing can also be applied to unequally spaced meshes.
For this problem the hardest part is to calculate the elements of the roughness matrix $\mathbf{D}$ for the given mesh, but this is beyond the scope of this paper.
%%%%%%%%%%%%%%%%%%%%%%%%%%%%%%%%%%%%%%% 
\section{Discussion}\label{sec:discussion}
%%%%%%%%%%%%%%%%%%%%%%%%%%%%%%%%%%%%%%%
\par %%%%%%%%%%%%%%%%%%%%%
In this paper we have introduced a weighted Whittaker--Henderson smoother with cross-validation.
The continually variable weights allow to optimally balance the spectral features against the noise.
Furthermore, Whittaker--Henderson smoothing does not use a moving window, or knots, to do smoothing.
This leads to less problems at the edges of the data and leads to smaller artifacts in higher derivatives of smoothed data.
Of course, our method has several limitations.
Firstly, our method is computationally relatively expensive, which deters its use for large data volumes.
The solution may lie in better minimization algorithms, faster linear solvers, or methods for describing the weights $w_i$ that do not rely on interpolating splines.
Secondly, there is a need for accurate noise variance to do error analysis.
Thirdly, there is always a risk of under- or over-adaptation of the weights.
This in turn leads to under- or oversmoothing respectively, and gives wrong credibility intervals. 
Therefore, it is vital to always verify results and to never blindly trust the results of any smoother.
\par %%%%%%%%%%%%%%%%%%%%%
For our synthetic data, we assumed independent, homoscedastic, normally distributed and stationary noise, which is the optimal case for linear (penalized) least-squares smoothers. 
On the contrary, real data can have outliers, require calibration, or have correlated, non-normal noise.
Outliers are a common problem and causes large errors if they are not taken care off.
If a way of detecting outliers is available, then we can, for example, use weighted Whittaker--Henderson smoothing to completely ignore outliers~\cite{Garcia2010CSDA}.
Whenever noise is correlated, two considerations must be taken.
Firstly, the bandwidth of the smoothed result is limited by the noise bandwidth.
This means that features past, or close to, the noise bandwidth will no longer be resolvable.
Secondly, the statistical tests, such as cross-validation, used to verify whether the smooth is correct have to take the noise correlations into account.
\par %%%%%%%%%%%%%%%%%%%%%
Whenever the noise is heteroscedastic or non-stationary, weighted Whittaker--Henderson smoothing should be used, as discussed in section~\ref{sec:weight_wh_smoothing}.
Whittaker--Henderson smoothing, like other linear (penalized) least-squares smoothers, works optimally for smoothing normally distributed noise.
Somewhat non-normally distributed errors are generally not a problem, as long as the errors have mean zero, but will lead to less-than-optimal smoothness.
In case that noise is highly non-normally distributed, such as for binomial data~\cite{Henderson1928PIMC}, it is necessary to use a different smoothing formalism.
Therefore, future work on a more robust smoother that takes these real features into account is highly beneficial.
%%%%%%%%%%%%%%%%%%%%%%%%%%%%%%%%%%%%%%% 
\section{Conclusions}\label{sec:conclusion}
%%%%%%%%%%%%%%%%%%%%%%%%%%%%%%%%%%%%%%%
\par %%%%%%%%%%%%%%%%%%%%%
In this article we study a frequently occurring challenge in experimental and numerical observations, namely, how to resolve characteristic features, such as spectral peaks, and derivatives from measured data with unavoidable noise. 
Therefore, we have developed a modified Whittaker--Henderson smoothing procedure that balances the spectral features and the noise.
In this surprisingly little-known procedure, our contribution is to introduce adjustable weights that are optimized using cross-validation. 
Using the measurement errors, a straightforward error analysis of the smoothed results is feasible.
\par %%%%%%%%%%%%%%%%%%%%%
As an example, we calculate the optical group delay dispersion of a Bragg reflector from synthetic phase data with noise to illustrate the effectiveness of the smoothing algorithm. 
The calculated credibility interval $\sigma_{\mathbf{z}}$ of the smoothed dispersion is on average almost \num{41} times smaller than the standard deviation of the noise without smoothing.
We find that \SI{68.9}{\percent} of all smoothed dispersion data points agree with the ground-truth dispersion within the $1\sigma_{\mathbf{z}}$ credibility interval.
Within the $2\sigma_{\mathbf{z}}$ credibility interval we find no less than \SI{95.8}{\percent} agreement.
Thus, the smoother faithfully reconstructs the group delay dispersion, allowing to observe details that otherwise remain buried in noise.
\par %%%%%%%%%%%%%%%%%%%%%
We recommend our weighted Whittaker--Henderson smoother with cross-validation whenever a single smoothing parameter does not suffice due to a large range of feature widths.
To further illustrate the power of our smoother, we demonstrate our smoother with several commonly occurring difficulties in data and data analysis.
We recommend our general framework of Whittaker--Henderson smoothing to properly smoothen unequally sampled data, and to obtain discontinuities, including discontinuous derivatives or kinks, and to properly smooth data in the vicinity of boundaries to the measurement domains.
%%%%%%%%%%%%%%%%%%%%%%%%%%%%%%%%%%%%%%%
\section*{acknowledgments}
%%%%%%%%%%%%%%%%%%%%%%%%%%%%%%%%%%%%%%%
We thank Lyuba Amitonova for stimulating discussions.
This work is supported by the Dutch Research Council NWO-TTW Perspectief program P21-20 ‘Optical coherence; optimal delivery and positioning’ (OPTIC) in collaboration with TUE, TUD, and ARCNL, and with industrial partners Anteryon, ASML, Demcon, JMO, Signify, and TNO. 
%%%%%%%%%%%%%%%%%%%%%%%%%%%%%%%%%%%%%%% 
\section*{Author Declarations}\label{sec:author_decl}
%%%%%%%%%%%%%%%%%%%%%%%%%%%%%%%%%%%%%%%
%%%%%%%%%%%%%%%%%%%%%%%%%% 
\subsection*{Conflict of Interest}\label{subsec:COI}
%%%%%%%%%%%%%%%%%%%%%%%%%%
The authors have no conflicts to disclose.
%%%%%%%%%%%%%%%%%%%%%%%%%% 
\subsection*{Author Contributions}\label{subsec:author_contr}
%%%%%%%%%%%%%%%%%%%%%%%%%%
\textbf{Bert Mulder}: Conceptualization; Data curation; Formal analysis; Investigation; Methodology; Project administration; Software; Visualization; Writing - original draft.
\textbf{Ad Lagendijk}: Supervision (supporting); Validation (supporting).
\textbf{Willem L. Vos}: Funding acquisition; Resources; Supervision (lead); Validation (lead); Writing - review \& editing.
%%%%%%%%%%%%%%%%%%%%%%%%%%%%%%%%%%%%%%% 
\vspace{.5cm}
\section*{Data Availability}\label{sec:DAS}
%%%%%%%%%%%%%%%%%%%%%%%%%%%%%%%%%%%%%%%
A collection of seven Python Notebooks containing code used for this publication is available~\cite{Mulder2026ZenodoWH} via the open-access repository Zenodo database that is developed under the European OpenAIRE program and operated by CERN~\cite{ZenodoDoi}. 
%%%%%%%%%%%%%%%%%%%%%%%%%%
\bibliography{whittaker.bib}
%%%%%%%%%%%%%%%%%%%%%%%%%%
\onecolumngrid
\clearpage
\appendix
%%%%%%%%%%%%%%%%%%%%%%%%%%%%%%%%%%%%%%% 
\section*{Appendix}\label{sec:appendix}
%%%%%%%%%%%%%%%%%%%%%%%%%%%%%%%%%%%%%%%
%%%%%%%%%%%%%%%%%%%%%%%%%%
\begin{figure*}[h!]
    \centering
    \includegraphics[width=\linewidth]{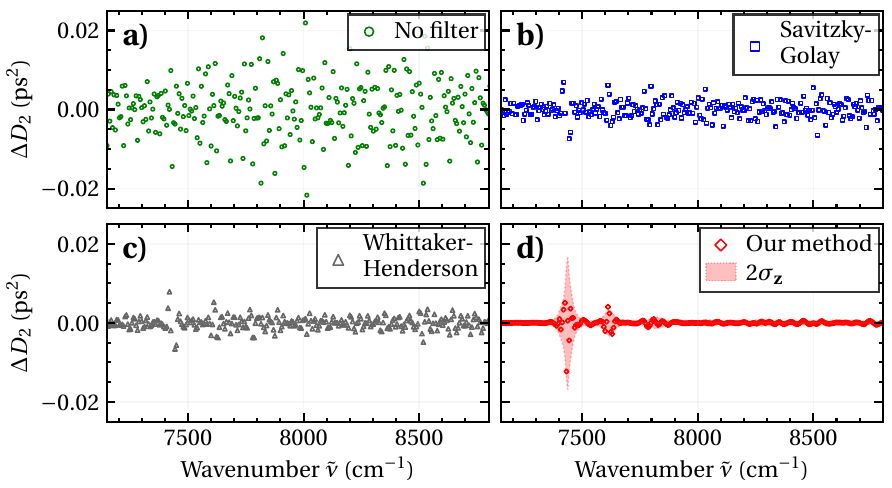}
    \caption{
        Residual of the calculated group delay dispersion $D_2(\omega)$ from noisy, synthetic reflectivity data, compared to the true dispersion without noise, of a Bragg reflector as shown in figure~\ref{fig:method_gdd_comparison}.
        \textbf{(a)} The second derivative needed for calculating dispersion is extremely sensitive to even small amounts of noise.
        The standard-deviation of the residual over all data without smoothing is $\sigma=\SI{6730}{\fs\squared}$.
        \textbf{(b)} The noisy reflectivity data is filtered by a fourth-order Savitzky--Golay filter with an optimal window length of nine points, before the dispersion is calculated.
        The standard-deviation of the residual over all data is reduced to $\sigma=\SI{3043}{\fs\squared}$, allowing to observe some features.
        \textbf{(c)} The noisy reflectivity data is smoothed by a fourth-order Whittaker--Henderson smoother with optimal smoothing parameter $\alpha=\num{0.110}$, before the dispersion is calculated.
        The standard-deviation of the residual over all data is further reduced to $\sigma=\SI{2499}{\fs\squared}$, owing to the better high-frequency noise suppression.
        \textbf{(d)} The noisy reflectivity data is reconstructed by our modified Whittaker--Henderson smoother, before the dispersion is calculated.
        The continuously variable smoothing weights allows strong noise reduction, yielding a standard-deviation of the residual over all data of only $\sigma=\SI{859}{\fs\squared}$.
        The shaded area denotes the estimated $2\sigma_{\mathbf{z}}$ credibility interval.
    }
    \label{fig:method_residual_comparison}
\end{figure*}
%%%%%%%%%%%%%%%%%%%%%%%%%%
%%%%%%%%%%%%%%%%%%%%%%%%%%
\end{document}